\documentclass[11pt]{article}
\usepackage{amsmath}
\usepackage{color}
\usepackage{graphics}
\usepackage{graphicx}
\usepackage{epsfig}
\usepackage{amssymb}
\usepackage{multirow}
\usepackage{float}

\usepackage{simplemargins} \setleftmargin{1in} \setrightmargin{1in}
\settopmargin{1in} \setbottommargin{1in}

\usepackage[]{lineno}

\newcommand{\bigU}{\mathbf{U}}

\newcommand{\XP}{\mathbf{X}_0}
\newcommand{\XS}{\mathbf{X}_1}
\newcommand{\GP}{\mathbf{\Gamma}_0}
\newcommand{\GS}{\mathbf{\Gamma}_1}
\newcommand{\GU}{\mathbf{\Gamma}_2}
\newcommand{\summa}{\frac{1}{l}\sum\limits_{k=0}^{l-1}}
\newcommand{\homega}{\mathbf{\hat{\Omega}}}
\newcommand{\hv}{\mathbf{\hat{V}}}
\newcommand{\hl}{\mathbf{\hat{\Lambda}}}

\let\oldequation\equation
\let\oldendequation\endequation

\renewenvironment{equation}
  {\linenomathNonumbers\oldequation}
  {\oldendequation\endlinenomath}

\begin{document}

\title{Application of the Non-Hermitian Singular Spectrum Analysis to the exponential retrieval problem}

\author{D.J Nicolsky$^1$ and G.S. Tipenko$^2$}
\maketitle

\date{~
\\
$^1$ Geophysical Institute, University of Alaska Fairbanks, \\ Fairbanks, PO Box 757320, Fairbanks, AK 99775, USA, djnicolsky@alaska.edu
\\
$^2$ Institute of Environmental Geoscience Russian Academy of Sciences,\\
13-2 Ulansky pereulok, PO Box 145, 101000 Moscow, Russia, gstipenko@mail.ru}

\begin{abstract}
We present a new approach to solve the exponential retrieval problem. We derive a stable technique, based on the singular value decomposition (\textsc{SVD}) of lag-covariance and cross-covariance matrices consisting of covariance coefficients computed for index translated copies of an initial time series. For these matrices a generalized eigenvalue problem is solved. The initial signal is mapped into the basis of the generalized eigenvectors and phase portraits are consequently analyzed. Pattern recognition techniques could be applied to distinguish phase portraits related to the exponentials and noise. Each frequency is evaluated by unwrapping phases of the corresponding portrait, detecting potential wrapping events and estimation of the phase slope. Efficiency of the proposed and existing methods is compared on the set of examples, including the white Gaussian and auto-regressive model noise\footnote{to appear in the Journal of the Russian Universities Radioelectronics.}.
\end{abstract}


\section*{Introduction}
The present paper originates from a classical problem in signal processing, namely: how to find a number of exponential constituents and their frequencies $\{\nu_{j}\}$ in a time series $\{f\left( k\right) \}_{k=0}^{m-1}$. One of the techniques is to assume that
\begin{equation}\label{eq:2}
f(k)=s(k)+\epsilon{w}(k),\ \ \ \ s(k)=\sum_{j=-n}^{n}c_{j}e^{i\nu
_{j}k}, \ \ \ \ k=0,1,\dots ,m-1
\end{equation}
and then to apply some least square method. Here, the complex-valued amplitudes $\{c_j\}$ and the real distinct frequencies $\{\nu _j\}$ are such that $c_{-j}=\bar{c}_j$, $\nu _{-j}=-\nu _{j}$. Note that $\nu_0=0$ and $c_0$ is a real-valued constant. The random component $w$ is commonly interpreted as noise, $s$ is called the signal and $\epsilon$ is a real constant.

A variety of subspace methods such as the Maximum Entropy Method \cite{Burg67}, MUltiple SIgnal Classification (MUSIC) \cite{Schmidt81}, Linear Prediction Methods \cite{Tufts82,Kay88}, Estimation of Signal Parameter via Rotational Invariance Technique (ESPRIT) \cite{Roy89}, Matrix Pencil (MP) \cite{Hua90}, and Minimum-Norm Method \cite{Kumaresan83} have been used to find the exponentials $\{\nu_j\}$ in the measured data $\{f(k)\}_{k=0}^{m{-}1}$. In \cite{Veen93}, a unification of concepts of the subspace methods is presented in terms of the singular value decomposition (SVD) \cite{Golub83} of the $d{\times}{l+1}$ trajectory matrix $\mathbf{X}$:
\begin{equation}\label{eq:defyk}
\mathbf{X}{=}(Y_0~ Y_1~{\dots}~Y_l), \ \ \ \
Y_k{=}(f(k{+}\kappa_1),f(k{+}\kappa_2),{\dots},f(k{+}\kappa_{d}))^t,
\end{equation}
for some constant $d$, translations $\kappa_i$, and the constant $l{=}m{-}\kappa_d$. We emphasize that in \cite{Veen93}, $\kappa_i{=}i{-}1$, whereas we propose to choose arbitrary translations. The new choice of translations allows us to increase the numerical rank \cite{Golub83} of the trajectory matrix $\mathbf{X}$, and to improve accuracy of frequency evaluation.

Our method together with ESPRIT and MP employ shift-invariance properties of the trajectory matrix $\mathbf{X}$, however, there are some differences. ESPRIT was developed to estimate the direction-of-arrival by a uniform linear array (ULA) of sensors. The data readings from the $i$-th sensor is associated with the $i$-th row of the trajectory matrix $\mathbf{X}$. The data in MP are similarly arranged in the row-wise format. Consequently, in both methods, the matrix $\mathbf{X}$ is partitioned into two submatrices $\mathbf{H}_0$ and $\mathbf{H}_1$ composed by the first $d{-}1$ and last $d{-}1$ rows of $\mathbf{X}$, respectively. Note that equal spacings between sensors in ULA yields $\kappa_i{=}i{-}1$ and hence the space shift-invariance property can be applied. If noise is absent and $d{=}2n+2$ then the space shift-invariance property let
us derive
\begin{equation}\label{eq:spaceinv}
\mathbf{H}_1=\mathbf{\Psi}\mathbf{H}_0,\ \ \
\mathbf{\Psi}=\{e^{i(j-1)\nu_k}\}\mathbf{\Lambda}\{e^{i(j-1)\nu_k}\}^{-1},
\
\mathbf{\Lambda}{=}\mbox{diag}[e^{i\nu_{{-}n}},\dots,e^{i\nu_{n}}],
\end{equation}
where $j=1,\dots,d-1$ and $k=-n,\dots,n$, and $\mathbf{\Lambda}$ is the eigenvalue matrix of $\mathbf{\Psi}$. For the non-uniform linear array (NULA) of sensors the translations $\{\kappa_i\}$ are arbitrary and hence the space shift-invariance in space property of $\mathbf{X}$ does not directly apply \cite{Buhren03}. However, if $d{=}2n{+}1$ then the matrix $\mathbf{X}$ has a time shift-invariance property:
\begin{equation}\label{eq:timeinv}
\mathbf{X}_1=\mathbf{\Omega}\mathbf{X}_0,\ \ \
\mathbf{\Omega}=\{e^{i\kappa_j\nu_k}\}\mathbf{\Lambda}\{e^{i\kappa_j\nu_k}\}^{-1},~~
\mathbf{\Lambda}{=}\mbox{diag}[e^{i\nu_{{-}n}},\dots,e^{i\nu_{n}}],
\end{equation}
where $j=1,\dots,d$ and $k=-n,\dots,n$; and the matrices $\mathbf{X}_0$, $\mathbf{X}_1$ are given by the first $l{-}1$ and last $l{-}1$ columns of the matrix $\mathbf{X}$, respectively; the matrix $\mathbf{\Lambda}$ is as in (\ref{eq:spaceinv}). The matrix
\[
\mathbf{V}=\{e^{i\kappa_j\nu_k}\}_{1{\leq}j{\leq}d,\ \ -n{\leq}k{\leq}n}
\]
is a generalized Vandermonde matrix and we assume that it is non-singular \cite{Vandermonde1,Vandermonde2}. Note that each frequency $\nu_j$ is given by the argument of the corresponding eigenvalues of $\mathbf{\Omega}$.

In the presence of noise, (\ref{eq:timeinv}) no longer holds. Therefore, we construct a $d{\times}d$ matrix
$\mathbf{\hat\Omega}$ such that $Y_k-\mathbf{\hat\Omega}Y_{k-1}$ is minimal in some sense, where $\{Y_k\}$ are given in (\ref{eq:defyk}). In the framework of perturbation analysis it is possible to show that if $d=2n+1$ then frequencies $\{\nu_j\}$ could be approximated by the arguments of the eigenvalues of $\mathbf{\hat\Omega}$. However, the number $2n+1$ of exponentials in the time series $f$ is \textit{a priori} unknown and needs to be found. To deal with this problem we propose a two step approach. In the first step, we select $d$ to be greater than the  number of exponentials found either by existing methods \cite{Akaike74,Schwartz78,Rissanen78,Wax85,Zhao86,Fuchs98,Badeau04} or by computing the rank of $\mathbf{X}_0$. We stress that we do not need to estimate the number of exponentials exactly at this step but to ensure that $d\ge2n+1$. In this case some eigenvalues of $\mathbf{\hat\Omega}$ are associated with exponentials, while others are related to the noise. Note that just taking into the account information about the eigenvalues, it is impossible to judge whether the eigenvalue of $\mathbf{\hat\Omega}$ is associated with the exponential or noise. Therefore, at the second step we cast the trajectory matrix $\XP$ into the basis of the eigenvectors of $\mathbf{\hat\Omega}$. In the new basis rows of $\XP$, that are associated with the exponentials, have a very specific structure, i.e. the phase portrait is either the circle or an arc. We hence propose to evaluate the number $n$ by a pattern recognition technique. We also show that the frequencies $\{\nu_j\}_{j=-n}^n$ estimated using the information carried by the rows are more accurate than those estimated by the eigenvalues of $\mathbf{\hat\Omega}$.

Before we proceed forward, we adopt notation for the inner product $(f,g)_l{=}{\summa}f(k)\overline{g(k)}$ and the norm $\|f\|_l{=}\sqrt{(f,f)_l}$ of time series $f{=}\{f(k)\}_{k=0}^{l-1}$ and $g{=}\{g(k)\}_{k=0}^{l-1}$. The operator $\overline{~\cdot~}$ stands for the complex conjugation, i.e. $\overline{a+ib}=a-ib$. If $l{=}\infty$, we define $\|f\|_\infty{=}\lim\limits_{l\to\infty}\|f\|_l$ and $(f,g)_\infty{=}\lim\limits_{l\to\infty}(f,g)_l$. Also, we defined the translation operator $\mathbf{T}$:
\[
\mathbf{T}{f}=(f(1),f(2),\dots,f(l-1)),  \ \ \
f=(f(0),f(1),\dots,f(l-2)),
\]
where $f{=}\{f(k)\}_{k=0}^{l-1}$ stands for a time series.


\section{The case of a single exponential}\label{sec:LinReg}
In this section, we consider a single exponential corrupted by noise:
\begin{equation}\label{eq:sinxp}
f(k)=s(k)+{\epsilon}w(k), \ \ \ \ s(k)=e^{i{\nu}k}
\end{equation}
and highlight key elements of the proposed technique. Our goal is to estimate the value of $\nu$ given values of $f(k)$ for $k=0,1,\dots,l$.

A number of methods have been developed to estimate a single exponential in the time series, e.g. \cite{Schmidt81,Tufts82,Roy89,Hua90,Kumaresan82,Tretter85,Stoica89,Kay89,Stoica97,Cedervall97,So05,Qian12,Li14,Zhao19}. Some of them are based on an observation that the exponential satisfies a first order auto-regressive process
\begin{equation}\label{eq:rotationidea}
s(k+1)=e^{i{\nu}k}s(k)
\end{equation}
and poles of the associated filter could be used to identify the frequency $\nu$ \cite{Stoica97,Li14}.

To estimate $\nu$ when the observations are corrupted by noise, it is possible to introduce some averaging by solving a first order autoregressive problem, i.e. finding the scalar value of $a$ such that the error term $e$ in the following relation is minimized:
\[
f(k+1)=af(k)+e(k+1),\ \ \ \ \ \ \|e\|_l\to\min,
\]
or in the matrix form:
\[
\XS = a\XP +\mathbf{E},\ \ \ \ \ \ \|\mathbf{E}\|\to\min,
\]
where $\XP=[f(0),f(1),{\dots},f(l{-}1)]$ and $\XS=[f(1),f(2),{\dots},f(l)]$ are complex-valued $1{\times}l{-}1$ matrices. The value of $a$ is found by solving the least squares and is given by $\hat\lambda_l$ such that
\[
\XS\XP^*=\hat\lambda_l\XP\XP^*,
\]
where ${}^*$ stands for the matrix conjugate transpose. Consequently, the frequency $\nu$ can be evaluated by $\tilde\nu_l=\Im(\ln(\hat\lambda_l))$, where $\hat\lambda_l$ is an eigenvalue in the case of $1\times1$ matrices $\XS\XP^*$ and $\XP\XP^*$. After some algebra, it is possible to derive that
\begin{equation} \label{eq:apeigen}
  \tilde\nu_l=\nu+\epsilon\Im\left((\mathbf{T}{w},\mathbf{T}{s})_l-(w,s)_l\right) + \epsilon^2\Im\left(e^{-i\nu}(\mathbf{T}{w},w)_l-\frac12\left[(s,w)_l+(\mathbf{T}{w},\mathbf{T}{s})_l\right]^2\right)+o(\epsilon^3).
\end{equation}

When information about a number of exponentials in the time series is missing, one might increase an order of the autoregressive model to find constants $a_1$ and $a_2$:
\[
f(k+1)=a_1 f(k)+a_2f(k-1)+e(k+1),\ \ \ \ \ \ \|e\|_l\to\min,
\]
or in the matrix form: find the matrix $\mathbf{A}$ such that
\begin{equation}\label{eqMatA}
\XS=\mathbf{A}\XP+\mathbf{E},\ \ \ \ \ \ \|\mathbf{E}\|_l\to\min,
\end{equation}
where
\[
\mathbf{A}=\left(
\begin{array}{cc}
  0 & 1 \\
  a_2 & a_1
\end{array}\right),
\]
$\XP=[Y_0,Y_1,{\dots},Y_{l{-}1}]$ and $\XS=[Y_1,Y_2,{\dots},Y_l]$ are complex-valued $2{\times}l{-}1$ matrices composed of the $2\times1$ information vectors ${Y}_k=[f(k), f(k+1)]^t$, and $\mathbf{E}$ stands for the $2{\times}l{-}1$ matrix associated with the noise. Note that in the case of the auto-regressive model of the first order, the information vectors $Y_k$ are scalars equal to $f(k)$.

Using least squares, the matrix $\mathbf{A}$ in (\ref{eqMatA}) is found by
\[
\mathbf{\hat{\Omega}}=\XS\XP^*(\XP\XP^*)^{-1}.
\]
In the case of the single infinite exponential corrupted by the white noise, i.e. $(w,w)_\infty=1$, $(s,w)_\infty=0$, and $(Tw,w)_\infty=0$ we obtain that $\hat{a}_1=e^{i\nu}/(\epsilon^2+2)$ and $\hat{a}_2=e^{2i\nu}/(\epsilon^2+2)$. Therefore the eigenvalues $\hat\lambda_k$ and eigenvectors $\mathbf{\hat{v}}_k$ of $\mathbf{\hat{\Omega}}$ are
\[
\hat\lambda_{1,2}=e^{i\nu}\frac{1\pm\sqrt{9+4\epsilon^2}}{2(2+\epsilon^2)}, \ \ \ \ \ \ \mathbf{\hat{v}}_{1,2}=\left[\begin{array}{c}
  -\hat\lambda_{2,1}  \\
  1
\end{array}\right].
\]
One may note that the argument of $\hat\lambda_1=e^{i\nu}(1-\epsilon^2/3)+O(\epsilon^4)$ could be used as an estimator of the frequency $\nu$. The other eigenvalue $\hat\lambda_2=-e^{i\nu}(1/2-\epsilon^2/12)+O(\epsilon^4)$ is rotated by the angle of $\pi$ with respect to the argument of $e^{i\nu}$. The arguments of eigenvalues may be thus used to estimate the frequency $\nu$, but $\hat\lambda_2$ provides a false estimate. The modulus of eigenvalues could be used to distinguish genuine and false estimates, e.g. eigenvalues with the absolute values significantly less than one could be associated with the damping exponentials and be discarded. At the same time, the eigenvectors $\mathbf{\hat{v}}_1$ and $\mathbf{\hat{v}}_2$ also carry the information about the exponentials.

In the proposed technique we look at the dynamics of trajectory matrix $\XP$ by mapping it to the basis of eigenvectors using the matrix $\mathbf{\hat{V}}=[\mathbf{\hat{v}}_1,\mathbf{\hat{v}}_2]$. An image of the information vector $Y_k$ in the new basis is
\[
Z_k=\mathbf{\hat{V}^{-1}}Y_k, \ \ \ \ \ k=0,\dots,l.
\]
In our case,
\[
\mathbf{\hat{V}^{-1}}=\frac1{2\sqrt{9+4\epsilon^2}}\left(
\begin{array}{cc}
   2e^{i\nu} & \sqrt{9+4\epsilon^2} + 1 \\
  -2e^{i\nu} & \sqrt{9+4\epsilon^2} - 1
\end{array}
\right)\approx\frac13\left(
\begin{array}{cc}
   e^{i\nu}(1-2\epsilon^2/9) & 2-\epsilon^2/9 \\
  -e^{i\nu}(1-2\epsilon^2/9) & 1+\epsilon^2/9
\end{array}
\right)+O(\epsilon^4),
\]
and hence
\[
Z_k=\left(\begin{array}{c}
   e^{ik\nu} \\
  0
\end{array}\right)+\frac\epsilon3
\left(\begin{array}{c}
  2w(k+1)+w(k)e^{ik\nu} \\
  w(k+1)-w(k)e^{ik\nu}
\end{array}\right)+\frac{\epsilon^2}9\left(\begin{array}{c}
   -e^{ik\nu} \\
   e^{ik\nu}
\end{array}\right)+O(\epsilon^3), \ \ \ \ \ k=0,\dots,l.
\]
The first coordinate of $Z_k$, i.e. $Z_{k,1}$ for $k=0,\dots,l$ rotates around the origin with an angle between $Z_{k+1,1}$ and $Z_{k,1}$ approximately equal to the frequency $\nu$. Its phase portrait $\{(\Re(Z_{k,1}),\Im(Z_{k,1}))\}_{k=0}^l$ resembles a unit circle (or an unit arc) with the center at the origin. The second coordinate of $Z_k$, i.e. $Z_{k,2}$ for $k=0,\dots,l$ does not have a particular well-defined behavior, i.e. its phase portrait is given by a set of points randomly centered around the origin. This difference in the phase portraits allows to distinguish pairs of the eigenvector-eigenvalue corresponding to the true frequency from their false counterparts.

We note that once the coordinate of $Z_k$ related to the exponential signal is established (in our case the first coordinate), then the problem is simplified. Any appropriate method of the frequency estimation could be applied to recover a single exponential in the time series.

\section{Linear regression approach in the case of multiple exponentials}
In this section, we extend our proposed approach to the time series composed of several exponentials and contaminated by noise. Now and for the rest of this article, we consider the information vectors $Y_k=(f(k+\kappa_1),f(k+\kappa_2),\dots,f(k+\kappa_{d}))^t$ associated with arbitrary translations $\{\kappa_i\}$. We consequently partition $Y_k$ into two time series $S_k$ and $W_k$:
\[
Y_k=S_k+{\epsilon}W_k, \ \ \
\]
where
$S_k=(s(k{+}\kappa_1),s(k{+}\kappa_2),{\dots},s(k{+}\kappa_{d}))^t$ and obtain that
\[
Y_k=\mathbf{V}\mathbf{\Lambda}^kC+{\epsilon}W_k, \ \ \ \ \ \ \ k=0,\dots,l.
\]
Here, $\mathbf{V}=\{e^{i\kappa_j\nu_k}\}_{1{\leq}j{\leq}d,~-n{\leq}k{\leq}n}$, $C=(c_{-n},\dots,c_{n})^t$, and $\mathbf{\Lambda}=\mbox{diag}[e^{i\nu_{{-}n}},\dots,e^{i\nu_{n}}]$.
Thus, the trajectory matrices $\XP$ and $\XS$ could be expressed as
\begin{eqnarray}
\XP&=&\mathbf{V}[C, \mathbf{\Lambda}C, \dots, \mathbf{\Lambda}^{l-1}C ]+\epsilon[W_0,W_1,\dots,W_{l-1}], \\
\XS&=&\mathbf{V}\mathbf{\Lambda}[C, \mathbf{\Lambda}C, \dots, \mathbf{\Lambda}^{l-1}C ]+\epsilon[W_1,W_2,\dots,W_l].
\end{eqnarray}
Note that each row of $[C, \mathbf{\Lambda}C, \dots, \mathbf{\Lambda}^{l-1}C ]$ is associated with a unique frequency. The phase portrait of the $k-$th row is given by the set of points $\{c_ke^{ij\nu_k}\}_{j=0}^{l-1}$, representing a circle/arc with the radius of $|c_k|$ around the origin. A key idea of the proposed method is to represent $\XP$ in a new basis, extract its rows and estimate a frequency associated with each phase portrait.

\subsection{Linear regression problem}
After some algebra, we derive that $Y_{k+1}$ and $Y_k$ are connected by the relation
\[
Y_{k+1}=\mathbf{\Omega}Y_k+\epsilon(W_{k+1}-\mathbf{\Omega}W_k), \ \ \ \ \ \mathbf{\Omega}=\mathbf{V}\mathbf{\Lambda}\mathbf{V}^+,
\]
where $\mathbf{V}^+=(\mathbf{V}^*\mathbf{V})^{-1}\mathbf{V}^*$ is the Moore-Penrose inverse of $\mathbf{V}$. In the case of $2n+1=d$, the matrix $\mathbf{V}$ is square, and hence $\mathbf{V}^+=\mathbf{V}^{-1}$. Furthermore, if  $2n+1=d$ then similar to (\ref{eq:rotationidea}) we obtain the relation between the signal components:
\[
S_{k+1}=\mathbf{\Omega}S_k, \ \
\mathbf{\Omega}=\mathbf{V}\mathbf{\Lambda}\mathbf{V}^{-1},
\]
and to find frequencies $\{\nu_k\}$ in the noiseless time series one needs to compute $\mathbf{\Omega}$ and then find its eigenvalues.

As in the case of the single exponential, we introduce averaging and approximate the matrix $\mathbf{\Omega}$ by the real-valued $d{\times}d$ matrix $\mathbf{\hat\Omega}$, which is the solution of the linear regression problem $\|Y_{k+1}-\mathbf{A}Y_k\|\to\min$. Namely, the matrix $\homega$ is called the best estimate of $\mathbf\Omega$, if $\homega=\arg\min_{\mathbf{A}}J(\mathbf{A})$, where
\begin{equation}\label{eq:11}
J(\mathbf{A})={\summa}E_k^*E_k, \ \ \ E_k=Y_{k+1}-\mathbf{A}Y_k, \
\ \ k=0,\dots,l-1.
\end{equation}
Note that in the case of multiples exponentials we do not know the number of exponentials contributing to the time series $f$. For now we assume that $2n+1\le d$. We discuss the selection of translations $\{\kappa_i\}_{i=1}^d$ and the dimensions of the information vectors $Y_k$ later in this section.

It is possible to prove, see Appendix \ref{ap:NoiseVariance} for the proofs, that if $\mbox{det}\GP{\neq}0$, then $J(\mathbf{A})$ has the minimum only at $\homega$, satisfying
\begin{equation}\label{lem12}
\GS{=}\homega\GP,
\end{equation}
and its minimal value is $J_{min}=\mathbf{tr}\{\GU{-}\GS\GP^{-1}\GS^{*}\}$, where
\[
\GP=\frac1l\mathbf{X}_0\mathbf{X}_0^*, \ \
\GS=\frac1l\mathbf{X}_1\mathbf{X}_0^*, \ \
\GU{=}\frac1l\XS\XS^*.
\]
If $2n+1=d$, then for the noiseless time series $f$, the solution $\mathbf{\hat\Omega}$ of the linear regression problem coincides with $\mathbf{\Omega}$, and for the noisy data the eigenvalues of $\mathbf{\Omega}$ and $\mathbf{\hat\Omega}$ are connected via
\begin{equation}\label{eq:eigvaapp}
\mathbf{\hat\Lambda}=\mathbf{\Lambda}+\sum_{n=1}^\infty\epsilon^n\mathbf{\Lambda}^{(n)},
\end{equation}
where $\mathbf{\Lambda}^{(n)}$ are some diagonal matrices. Hence, we can use the eigenvalues of $\mathbf{\hat\Omega}=\GS\GP^{-1}$ to evaluate those corresponding to $\mathbf{\Omega}$ and by computing the argument of eigenvalues to find the frequencies.

There are some difficulties since the number of exponentials is \textit{a priori} unknown, but it could be estimated by methods discussed in \cite{Akaike74,Schwartz78,Rissanen78,Wax85,Zhao86,Fuchs98,Badeau04} or by computing the rank of $\mathbf{X}_0$. In the case of noiseless time series, we have $\mbox{rank}(\mathbf{X}_0)=2n+1$ for values $d\ge2n+1$. This means that after a certain increase in $d$, the rank of $\mathbf{X}_0$ stays constant and this threshold could be used to estimate the number of exponentials. In case of the noisy data, one can instead estimate the numerical rank of $\GP$ by SVD \cite{Veen93}. Under the white noise assumption and $m\to\infty$, the singular values $\{\lambda_k\}_{k=1}^d$ of $\GP$ according to \cite{Veen93,Li14} are such that
\[
\lambda_k\approx\left\{
\begin{aligned}
  \mu_k+\sigma^2, & ~~~ k=1,\dots,2n+1\\
  \sigma^2, & ~~~k=2n+2,\dots,d
\end{aligned}\right.,
\]
where $\mu_k>0$ are the singular values associated with the signal component and $\sigma^2$ stands for the noise variance. The values of $\mu_k$ depend on the choice of translations $\{\kappa_i\}_{i=1}^d$ and could be close to zero if the information vectors $Y_i$ are almost linearly dependent. By varying the translations, it is possible to make the information vectors more linearly interdependent, consequently increase $\mu_k$, and thus to obtain a notable distinctions between the singular values associated with the signal and noise. Note that the singular values associated with the white noise are constant. In the case when $m$ is not large or when the noise is not white, smallest singular values $\mu_k$ could start to overlap with those related to noise \cite{Stewart91,Li92} and hence the notable decrease may be missing. Nevertheless, the SVD provides a good estimate for the numerical rank of matrix \cite{Veen93}.

We note that if $d>2n+1$ then some eigenvalues of $\mathbf{\hat\Omega}$ approximate the frequencies $\{\nu_i\}$ while the other are associated with the noise. As in the case with the single exponential, selection of eigenvalues associated with the exponentials is a matter of belief if no information regarding the noise structure is provided. On other hand, the eigenvectors of $\homega$ can bring more information to decide whether the eigenvalue-eigenvector pair is associated with the exponential or noise. The ideas of using the eigenvectors are closely related to the time series decomposition by the Singular Spectrum Analysis (SSA) \cite{Broomhead86,Vautard89}.

\subsection{Time series decomposition and the principal component approach}\label{NHSSADecomp}
Before introducing the principal components using eigenvectors of $\mathbf{\hat\Omega}$, we briefly review some key points of the Singular Spectrum Analysis (SSA). Our approach exploits ideas of SSA, yet SSA and its various modifications such as Monte Carlo SSA \cite{Ghil91,Allen92a,Allen92b} and Multiscale SSA \cite{Yiou00}, Random Lag SSA \cite{Varadi99}, Oblique SSA \cite{Golyandina14} do not compute the frequencies $\{\nu\}$ but allow representation of the data $f$ in a new convenient way. In particular, SSA relies on the Karhunen-Lo\`eve decomposition of the correlation matrix
\[
\mathbf{\Gamma}_0=\frac1l\mathbf{U}\mathbf{\Sigma^2}\mathbf{U}^*,
\]
and on representation of the vectors $\{Y_k\}$ in a coordinate system defined by the eigenvectors $\{\mathbf{u}_k\}$ of $\mathbf{\Gamma}_0$, or namely

\[
\mathbf{P}=\mathbf{U}^*\mathbf{X}_0, \ \ \ \
\mathbf{U}=\{\mathbf{u}_1,\dots,\mathbf{u}_d\},
\]
where the rows $\{\mathbf{p}_k\}$ of the matrix $\mathbf{P}$ can be seen as the coordinates of $\{Y_k\}$ in the orthogonal base $\{\mathbf{u}_k\}$ and are commonly called principal components. Note that the vectors $\{\mathbf{p}_k\}$ have the important property of orthogonality $\mathbf{PP}^*=\mathbf{\Sigma}^2$. However, arbitrary exponentials do not have to be orthogonal with respect to the inner product $(\cdot,\cdot)_l$ due to the finite number of sampling points $k=0,\dots,l-1$. Therefore, each principal component $\mathbf{p}_k$ is usually a linear combination
of exponentials even for the noiseless data. We would like to emphasize that even for the noiseless data, there is no one-to-one correspondence between the exponentials and the principal components $\mathbf{p}_k$ \cite{Goljandina01}. Hence, our goal is to obtain a one-to-one correspondence between the frequencies $\{\nu\}$ and certain objects in the absence of noise as follows.

We achieve our goal by representing the information vectors $Y_k$ in the basis of eigenvectors of $\homega$ instead of the basis associated with $\GP$ as it completed in SSA. Consequently we call the proposed method the Non-Hermitian Singular Spectrum Analysis (NH-SSA). For the rest of this article we assume that the eigenvalues of $\homega$ are all different and the Jordan decomposition of $\homega$ holds as
\begin{equation}\label{eq:41}
\homega=\hv\hl\hv^{-1}.
\end{equation}
We define a central object in our approach, namely the image of information vectors $Y_k$ in the basis of $\hv$:
\begin{equation}\label{eq:46}
Z_k=\hv^{-1}Y_k, \ \ \ \ \ k=0,{\dots},l.
\end{equation}

Using the small perturbation theory it is possible to show that, in the case of $d=2n+1$, if all eigenvalues of $\mathbf{\Omega}$ are distinct then the eigenvector matrix $\mathbf{\hat{V}}$ is
such that
\begin{equation}\label{lem21}
\mathbf{\hat{V}}^{-1}=(\mathbf{I}+\epsilon\mathbf{R})\mathbf{V}^{-1}.
\end{equation}
Here, the matrix $\mathbf{R}(\epsilon)$ is analytic with respect to $\epsilon$ and
$\mathbf{R}(\epsilon)=\sum_{n=0}^\infty\epsilon^n\mathbf{R}^{(n+1)}$, where each diagonal entry of $\mathbf{R}^{(n+1)}$ vanishes for any $n$. Recalling that $Z_k=\hv^{-1}Y_k$, we have
\begin{equation}\label{eq:zkdef}
Z_k=\mathbf{\Lambda^k}C+\epsilon\big[\mathbf{R}\mathbf{\Lambda^k}C+\mathbf{\hat{V}}^{-1}W_k\big],
\end{equation}
or in its coordinate-wisely form as
\begin{equation}\label{eq:pcdef}
Z_{k+1,j}=\lambda_j^kc_j+\epsilon\eta_{k,j}, \ \
\eta_{k,j}=\sum_{i=1}^{2n+1}(R_{ji}\lambda^k_ic_i+{V_{ji}^{-1}W_{k,i}}),
\ \ \ \ j=1,\dots,2n+1.
\end{equation}
Here, $j$ stands for the row index of the vector $Z_k{=}(Z_{k,1},\dots,Z_{k,2n{+}1})^t$, $\lambda_j=e^{i\nu_j}$ is
the eigenvalue of $\mathbf{\Omega}$, and $\eta_{k,j}$ represents noise. For each value of $j$, the consecutive values of $Z_{k,j}$, $k=0,\dots,l$ rotate around the origin, each time turning by the angle of $\nu_j$ (up to the level of noise $O(\epsilon)$). Furthermore, if $d=2n+1$ and $\epsilon\to0$ then there is one-to-one correspondence between the rows of $Z_k$ and exponentials. 


In many practical applications, a number $2n+1$ of exponentials in the time series $f$ is unknown. Therefore, in order not to loose some exponentials, we need to have $d\ge2n+1$. This implies that we decompose the noise $w$ into exponentials. Therefore, some rows of $Z_k$ are attributed to noise and do not have a stable ``rotational'' pattern around the origin, as was discussed in Section 1 for the case of the single exponential.

\begin{figure}[ptb]
\centering
\includegraphics[width=0.94\columnwidth]{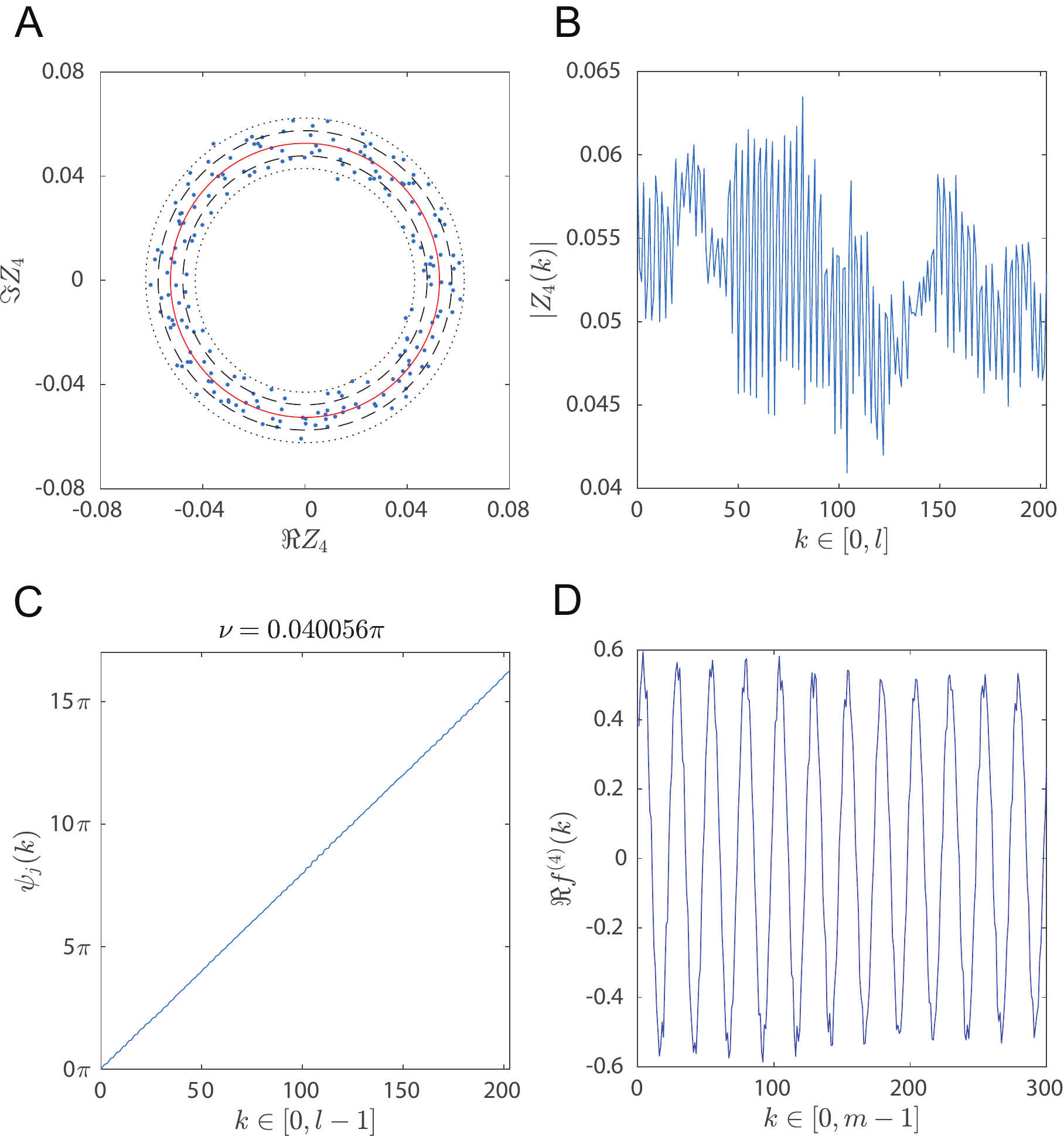}\caption{Typical results for the sequence $\{Z_{k,j}\}_{k=0}^l$ associated with an exponential. (A) The phase portrait $\{(\Re(Z_{k,j}),\Im(Z_{k,j}))\}_{k=0}^l$ lies between two concentric circles; dashed and dotted lines represent one and two standard deviations from the mean modulus; (B) The modulus of $\{Z_{k,j}\}_{k=0}^l$ is bounded from zero; (C) The phase $\psi_j(k)$ is a linearly increasing function; (D) Mapping of $\{Z_{k,j}\}_{k=0}^l$ back to the original space $f^{(j)}(k)$ is an exponential.}\label{fig:rotation4}
\end{figure}

\begin{figure}[ptb]
\centering
\includegraphics[width=0.94\columnwidth]{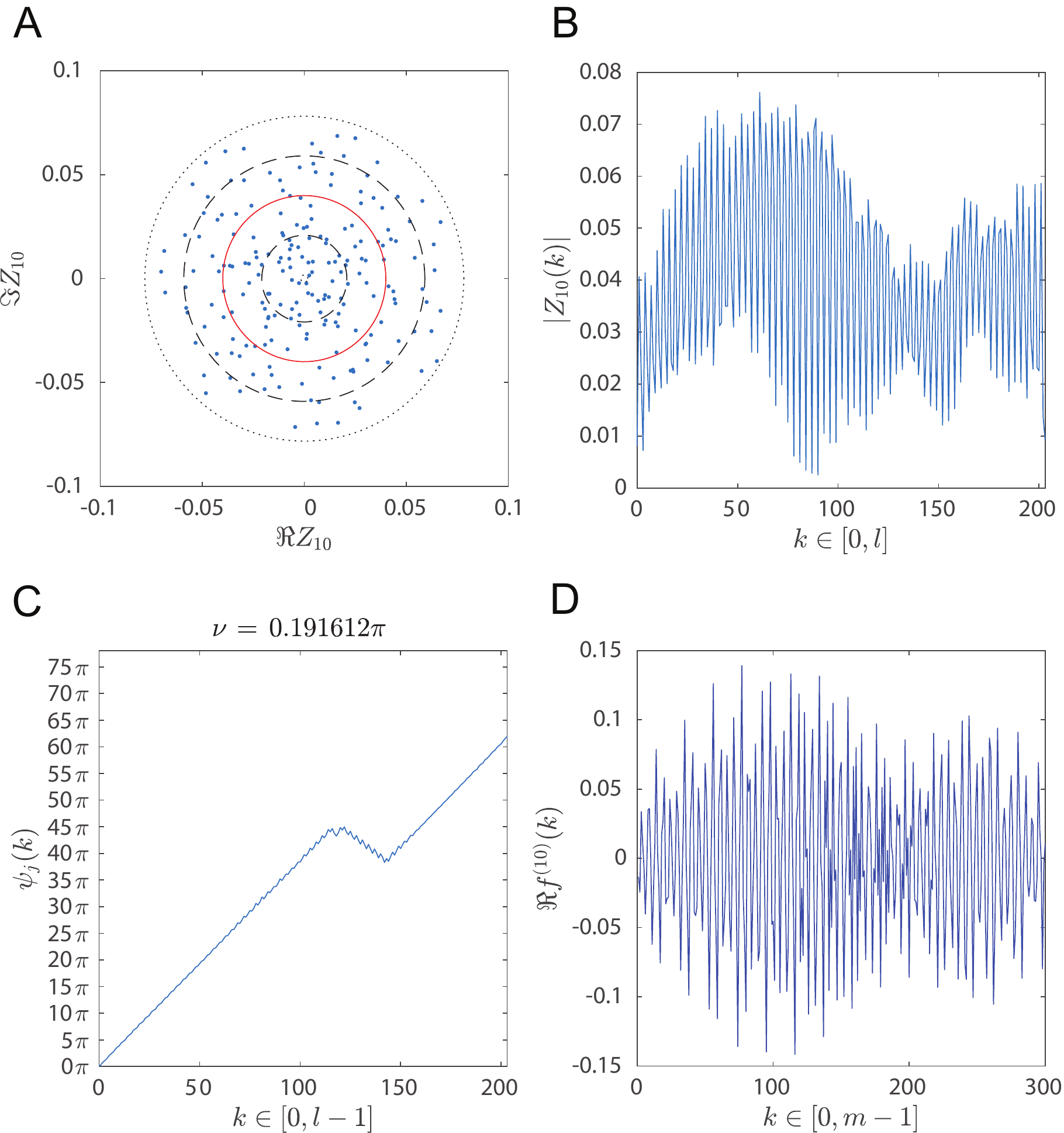}\caption{Typical results for the sequence $\{Z_{k,j}\}_{k=0}^l$ associated with noise. (A) The phase portrait $\{(\Re(Z_{k,j}),\Im(Z_{k,j}))\}_{k=0}^l$ is randomly distributed around the origin; (B) The modulus of $\{Z_{k,j}\}_{k=0}^l$ is not bounded from zero; (C) The phase $\psi_j(k)$ can have one or multiple ``wrapping'' events \cite{Fowler02}; (D) Mapping of $\{Z_{k,j}\}_{k=0}^l$ back to the original space $f^{(j)}(k)$ is an irregular signal. }\label{fig:rotation10}
\end{figure}

Thus, to decide whether the $j$-th row of $Z_k$ is associated with the noise or exponentials, we apply the pattern recognition technique by visualizing dynamics of $\{Z_{k,j}\}_{k=0}^l$. Namely, the sequence $\{Z_{k,j}\}_{k=0}^l$ is thought to represent a single exponential if the phase portrait, i.e. the set of points $\{(\Re(Z_{k,j}),\Im(Z_{k,j}))\}_{k=0}^l$ lies between two concentric circles, see Figure \ref{fig:rotation4}a; the modulus $|Z_{k,j}|$ is bounded around a certain constant. One other hand, the sequence $\{Z_{k,j}\}_{k=0}^l$ is associated with noise when the phase portrait $\{(\Re(Z_{k,j}),\Im(Z_{k,j}))\}_{k=0}^l$ is randomly distributed around the center, as shown in Figure \ref{fig:rotation10}a; the modulus $|Z_{k,j}|$ significantly varies. Furthermore, for the single exponential signal, the phase $\psi_j(k)$ is linearly increasing as shown in Figure \ref{fig:rotation4}c, while for $\{Z_{k,j}\}_{k=0}^l$ related to noise the phase $\psi_j(m)=\sum_{k=0}^{m-1}\Im\left(\ln\left({Z_{k+1,j}}/{Z_{k,j}}\right)\right)$ could have some irregularities, see Figure \ref{fig:rotation10}c.

Once the rows of $Z_k$ associated with the exponentials are established the process is greatly simplified, note that each $\{Z_{k,j}\}_{k=0}^l$ represents a single exponential and a whole variety of the frequency estimation methods could be applied to recover the frequency.

Finally, we note that similar to the SSA methodology, sequences $\{Z_{k,j}\}_{k=0}^l$ associated with either the exponentials or noise could be grouped together and mapped back to the original space as follows. For example, mapping of $\{Z_{k,j}\}_{k=0}^l$ associated with a single exponential back to the original space results in the trajectory matrix $\mathbf{\tilde{X}}_0^{(j)}=\{c_je^{i(\kappa_s+k)\nu_j}+O(\epsilon)\}_{ks}$, which is a Hankel matrix when the translations $\kappa_s=s-1$. We average elements in $\mathbf{\tilde{X}}_0^{(j)}$ and thus map the sequence $\{Z_{k,j}\}_{k=0}^l$ to the original space as $f^{(j)}(k)=c_je^{i\nu_jk}+O(\epsilon)$, $k=0,\dots,m-1$. Figures \ref{fig:rotation4}d and \ref{fig:rotation10}d show mappings of $\{Z_{k,j}\}_{k=0}^l$ associated with the exponential and noise, respectively, to the original space. Note that $\Re{f}^{(j)}(k)$ associated with the exponential is a cosine with an approximately constant amplitude, while the mapping of $\{Z_{k,j}\}_{k=0}^l$ related to the noise results in some irregular structure. Similar to SSA, sequences $\{Z_{k,j}\}_{k=0}^l$ associated with noise could be all grouped together and mapped back to $\hat{w}$. We thus produce a decomposition of the original signal $f=\hat{f}+\hat{w}+O(\epsilon)$, where $\hat{f}=\sum_jf^{(j)}$.

Decomposition of the time series $f$ into a sum of $f^{(j)}$, where each $f^{(j)}$ is associated with the single exponential, depends on the noise variance, proposed number $n$ of exponentials and translations $\{\kappa\}$. In the current model we use translations such that $\kappa_i{=}(i{-}1)\bar{m}$, where $\bar{m}$ is called the multiplicity. Hence, the decomposition depends on the pair of parameters $(d,\bar{m})$. The multiplicity $\bar{m}$ can be chosen arbitrarily. However, our numerical experiments show that $\bar{m}$ should be of the same order as the expected period. In this case, the information vectors $\{Y_k\}$ tend to be more linearly independent in the computational sense and the matrices $\GP$ and $\GS$ be better conditioned.

\section{Numerical realization}
In this section, we provide a numerical algorithm to evaluate $\mathbf{\hat{V}}^{-1}$ in order to map the time series into the principal components. Please note that since $\mathbf{\Gamma}_0$ and $\mathbf{\Gamma}_1$ are typically ill-conditioned matrices and the straightforward calculation of eigenvectors $\mathbf{V}$ of $\mathbf{\hat\Omega}=\mathbf{\Gamma}_1\mathbf{\Gamma}_0^{-1}$ could lead to computational errors \cite{Golub99}. We propose to maximize the ``numerical'' rank of $\mathbf{\Gamma}_0$ and $\mathbf{\Gamma}_1$  by varying $\bar{m}$. Further, noting that the SVD is less error prone for the symmetrical positive-definite matrices than the eigenvalue decomposition, in order to achieve a better numerical accuracy we suggest to exploit the SVD of $\mathbf{X}_0$ and $\mathbf{X}_1$ as follows.

Assuming that trajectory matrices $\XP,\XS$ have the rank equal to $2n{+}1$ we denote the SVD of $\XP$ and of $\XS $ as
\begin{equation}\label{eq:svd1}
\XP=\mathbf{U}_{0}\mathbf{D}_{0}\mathbf{\Pi}{\mathbf{W}_{0}^{*}},\
\ \
\XS=\mathbf{U}_{1}\mathbf{D}_{1}\mathbf{\Pi}{\mathbf{W}_{1}^{*}},
\end{equation}
where $2n{+}1{\times}l$ matrix $\mathbf{\Pi}$ stands for the projector mapping of $\mathbb{R}^{l}$ onto $\mathbb{R}^{(2n{+}1)}$, and where the unitary matrices $\mathbf{U}_s,\mathbf{W}_s$ and the non-negative diagonal matrix $\mathbf{D}_s$ are of size $2n{+}1{\times}2n{+}1$, $l{\times}l$ and $2n{+}1{\times}2n{+}1$, respectively for $s=0,1$. From (\ref{eq:svd1}) it follows that the equation $\GS={\homega}\GP $ is equivalent to
\begin{equation}
\mathbf{Q}=\mathbf{\Upsilon{R}},
\end{equation}
where
$\mathbf{Q}=\mathbf{\Pi}{\mathbf{W}_{1}^{*}}{\mathbf{W}_{0}}\mathbf{\Pi}^{*}$,
$\mathbf{\Upsilon}=\mathbf{D}_{1}^{-1}\mathbf{U}_{1}^{*}{\homega}\mathbf{U}_{1}\mathbf{D}_{1}$, $\mathbf{R}=\mathbf{D}_{1}^{-1}\mathbf{U}_{1}^{*}\mathbf{U}_{0}\mathbf{D}_{0}$. Let us emphasize that the matrix $\mathbf{Q}$ is a projection of the unitary matrix ${\mathbf{W}_{1}^{*}}{\mathbf{W}_{0}}$, and therefore $\|\mathbf{Q}\|\leq {1}$. The matrix $\mathbf{\Upsilon}$ has the same spectrum $\hl$ as ${\homega}$ has.

Let us denote by $\mathbf{\Phi}$ the matrix consisting of the generalized eigenvectors of the matrices
$(\mathbf{Q},\mathbf{R})$, namely $\mathbf{Q}{\mathbf{\Phi}}=\mathbf{R}\mathbf{\Phi}\hl$. Taking into the account that
$\mathbf{Q}=\mathbf{R}\mathbf{\Phi}\hl{\mathbf{\Phi}}^{-1}=\mathbf{\Upsilon}\mathbf{R}$ and the definition of the matrix $\mathbf{\Upsilon}$ it is easy to check that
\begin{equation}\label{eq:42}
\homega=(\mathbf{U}_{1}\mathbf{D}_{1}\mathbf{R}\mathbf{\Phi})\hl(\mathbf{U}_{1}\mathbf{D}_{1}\mathbf{R}\mathbf{\Phi})^{-1}.
\end{equation}
From the Jordan decomposition of $\homega$ it follows that the eigenvectors $\hv$ of $\homega$ and generalized
eigenvectors $\mathbf\Phi$ are connected by
\begin{equation}\label{eq:45}
\hv=\mathbf{U}_{1}\mathbf{D}_{1}\mathbf{R}\mathbf{\Phi}=\mathbf{U}_{0}\mathbf{D}_{0}\mathbf{\Phi}.
\end{equation}
Therefore, by definition (\ref{eq:46}) we obtain
\begin{equation}\label{eq:46dd}
Z_k=\hv^{-1}Y_k=\mathbf{\Phi}^{-1}\mathbf{D}_{0}^{-1}\mathbf{U}_0^{*}Y_k,
~~~~~k=1,{\dots},l.
\end{equation}

Finally, we combine all steps together and provide the numerical algorithm to estimate exponentials in the time series:

\textbf{Input:} the time series $f(k), k=0,1,\dots,m-1$.
\begin{enumerate}
\item Pick a value of $d>n$ and consider a set of multiplicities $\bar{m}=0,1,2,\dots$
\item For a selected value of $\bar{m}$, compute the translations $\kappa_k=(k-1)\bar{m}$, where $k=0,1,\dots,d$
\item Form the information vectors $\{Y_k\}_{k=0}^l$ and trajectory matrices $\XP$ and $\XS$
\item Repeat Steps 1-3 while computing the condition number $\textrm{cond}(\mathbf{X}_0)=\|\mathbf{X}_0\|\|\mathbf{X}_0^{-1}\|$ on a grid of $(d,\bar{m})$
\item Select a pair of $(d,\bar{m})$, where $\textrm{cond}(\mathbf{X}_0)$ attains the minimum and compute trajectory matrices $\XP$ and $\XS$ for the found pair of $d$ and $\bar{m}$
\item Perform the SVD on matrices: $\XP=\mathbf{U}_{0}\mathbf{D}_{0}\mathbf{\Pi}{\mathbf{W}_{0}^{*}},~~~
\XS=\mathbf{U}_{1}\mathbf{D}_{1}\mathbf{\Pi}{\mathbf{W}_{1}^{*}}$

\item Calculate $\mathbf{Q}=\mathbf{\Pi}{\mathbf{W}_{1}^{*}}{\mathbf{W}_{0}}\mathbf{\Pi}^{*}$ and
$\mathbf{R}=\mathbf{D}_{1}^{-1}\mathbf{U}_{1}^{*}\mathbf{U}_{0}\mathbf{D}_{0}$
\item Evaluate the generalized eigenvalues $\hl=\{\hat\lambda_0,\hat\lambda_1,\dots,\hat\lambda_l\}$ and eigenvectors $\mathbf{\Phi}$ for the pair of matrices $\mathbf{Q}$ and $\mathbf{R}$ such that $\mathbf{Q}\mathbf{\Phi}=\mathbf{R}\mathbf{\Phi}\hl$
\item Compute map information vectors $Y_k$ into the new basis: $Z_k=\mathbf{\Phi}^{-1}\mathbf{D}_{0}^{-1}\mathbf{U}_0^{*}Y_k$, $k=0,\dots,l$
\item Discard rows $\{Z_{k,j}\}_{k=0}^l$ associated with eigenvalues $\lambda_j$ such that $|\hat\lambda_j|<\lambda_c$, where $\lambda_c$ is a given threshold. These rows are related to the signal components $f^{(j)}$ damping with time. The value of $\lambda_c$ could be chosen such that a number of generalized eigenvalues $|\hat\lambda_j|<\lambda_c$ is the same as the number of singular values of $\GP$ associated with the signal.
\item Apply pattern recognition technique to the remaining $\{Z_{k,j}\}_{k=0}^l$ and evaluate the number $n$ of exponentials. The number $n$ of exponentials is determined by the number of $\{Z_{k,j}\}_{k=0}^l$ which have the graph lying in a vicinity of the unit circle in the complex plane.
\item Apply one of the frequency estimation methods to recover a single exponential in $\{Z_{k,j}\}_{k=0}^l$ associated with the regular patterns. Here, we use ESPRIT to find a single exponential in $\{Z_{k,j}\}_{k=0}^l$.
\end{enumerate}
\textbf{Output:} List the frequencies $\{\nu_j\}$ associated with $\{Z_{k,j}\}_{k=0}^l$ showing the regular patterns.

\section{Comparison with other high resolution methods}
\subsection{Signal corrupted by the white noise}
Let us consider the signal $s$ consisting of four unit-amplitude cosines sampled at $m=300$ points. The frequencies for these cosines are $\nu_1/2\pi=0.04$, $\nu_2=0.06$, $\nu_3=0.07$, and $\nu_4=0.12$. The signal $s$ is corrupted by the white Gaussian noise $w$ with the zero mean $\textrm{E}(w)=0$ and variance $\textrm{Var}(w)=1$. In this case, the SNR$_{\textrm{dB}}$ defined by $10\log_{10}\left(\|s\|_m^2/\|w\|_m^2\right)$ is approximately 3.5 dB.

\begin{figure}[H]
\centering
\includegraphics[width=.7\columnwidth]{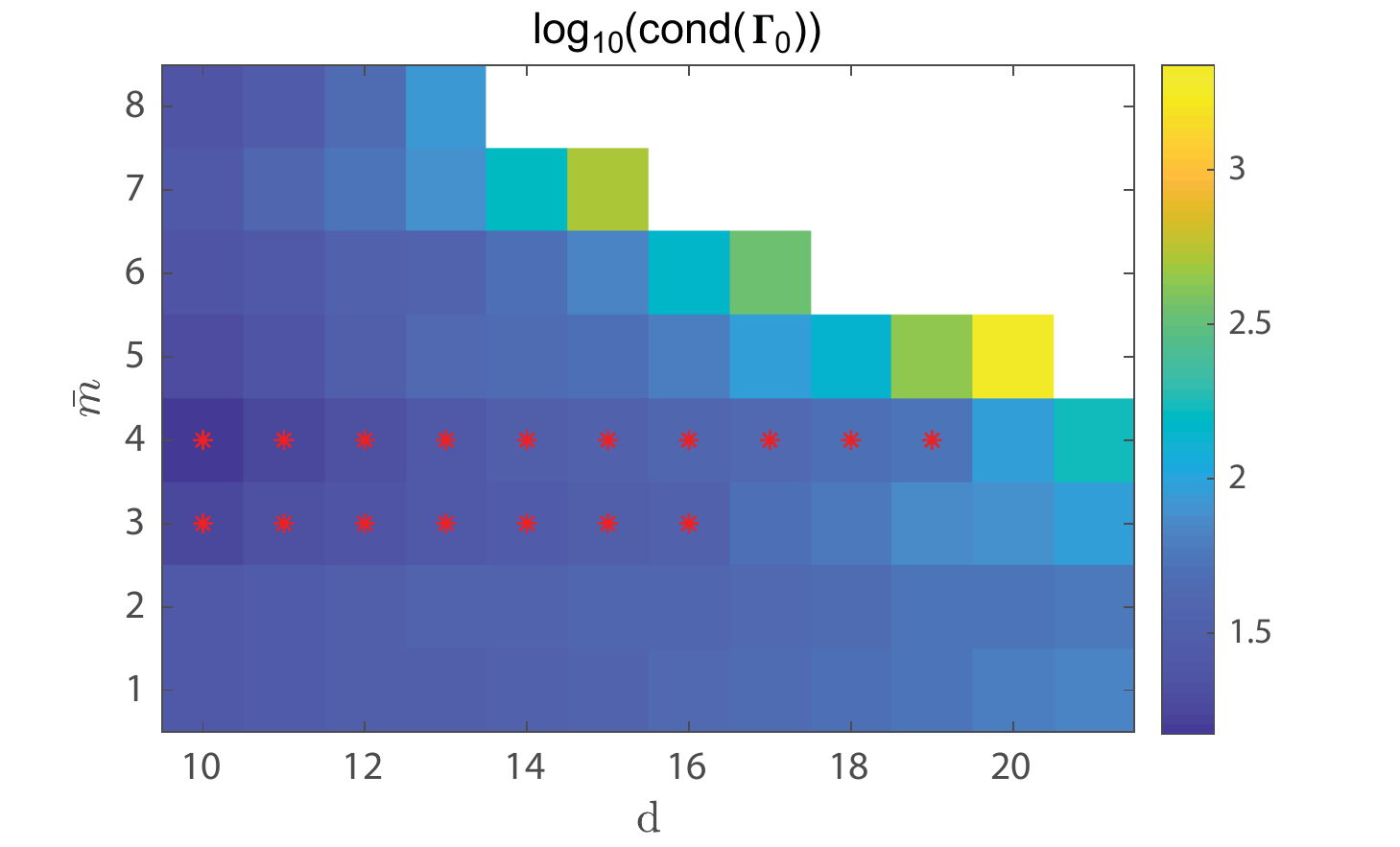}
\caption{Dependence of $\textrm{cond}(\GP)$ on the multiplicity $\bar{m}$ and the size of $\GP$. Cells marked by asterisks, where the condition number is minimal, are potential candidates to choose a pair of $(\bar{m},d)$.}\label{fig:cond}
\end{figure}

For a given realization of the time series, we compute the condition number $\textrm{cond}(\GP)$ on the grid of $(d,\bar{m})$, as shown in Figure \ref{fig:cond}. There are several potential pairs of $\bar{m}$ and $d$, marked by red asterisks, where the condition number is close to its minimum. We select the value of $\bar{m}$ equal to 4, while the choice for $d$ is less restrictive, at the same time is better to select $d$ rather large in order to increase the size of $\GP$ and reduce influence of noise on the information-carrying components.

\begin{figure}[H]
\centering
\includegraphics[width=.85\columnwidth]{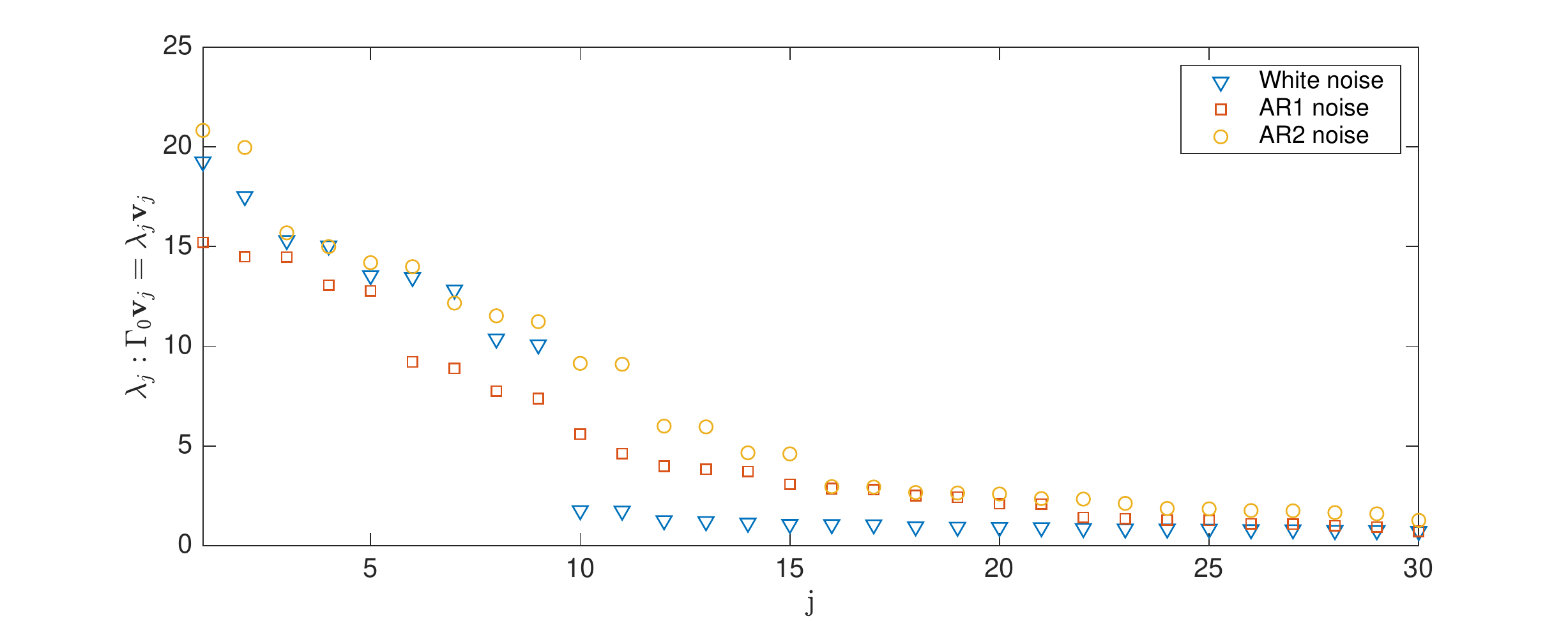}
\caption{Eigenvalues of the auto-covariance matrix $\GP$ for the signal consisting of a sum of eight exponentials and the constant. The signal is corrupted by either the white Gaussian noise or auto-regressive noise of the first and second orders.}\label{fig:eig}
\end{figure}

Our experience reveals that it is important to select the value of $d$ to be at least three to four times larger than the number of exponentials, which could be estimated, for example, by analyzing eigenvalues of the auto-covariance matrix $\GP$, shown by blue triangles in Figure \ref{fig:eig}. We note that a significant drop occurs at the $9^{th}$ eigenvalue. Hence, $\bar{n}=4$ (since the cosines are used to define the signal). Therefore, while applying the NHSSA method, we assume $d=18$, however any number greater than 12 will be also sufficient. We stress that the precise determination of $d$ is not important. Additional dimensions $d-n$ are used to decompose the noise into some exponentials and these false exponentials are discarded later at either Step 10 or 11.

\begin{figure}[H]
\centering
\includegraphics[width=.85\columnwidth]{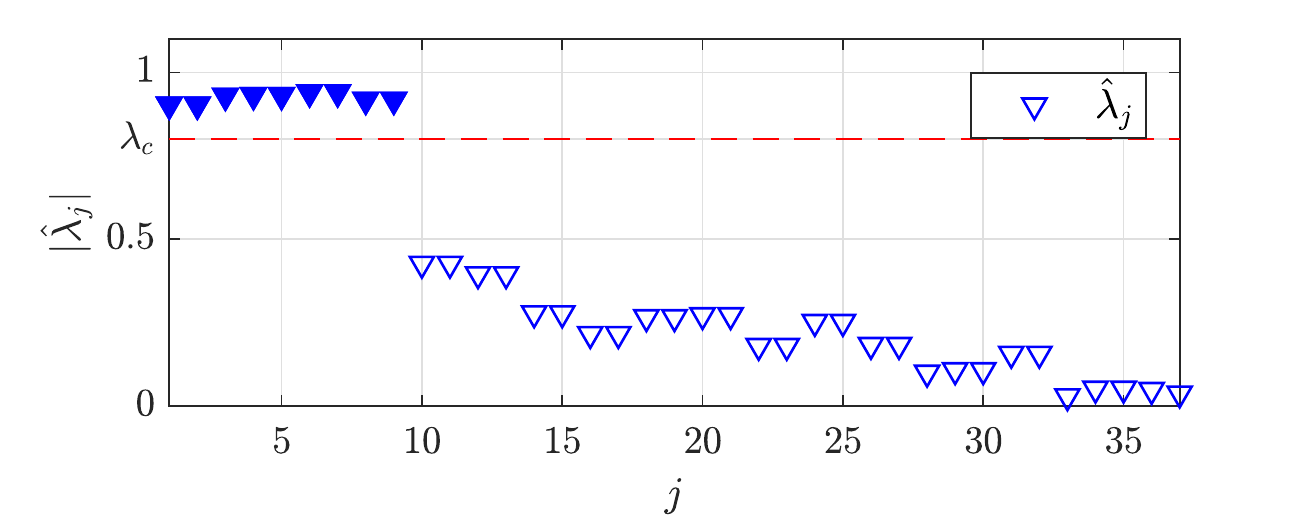}
\caption{Generalized eigenvalues $\hl=\{\hat\lambda_0,\hat\lambda_1,\dots,\hat\lambda_l\}$ for the pencil $\mathbf{Q}\mathbf{\Phi}=\mathbf{R}\mathbf{\Phi}\hl$.}\label{fig:eig_lambda}
\end{figure}

A number of these false estimates is controlled by the threshold parameter $\lambda_c$ in Step 10. If the value of $\lambda_c$ is reduced then more false estimates are recovered, but then could be discarded at Step 11. At the same time, if the value of $\lambda_c$ is increased then NHSSA may start to omit true frequencies. The value of $\lambda_c$ could be chosen such that a number of generalized eigenvalues $|\hat\lambda_j|<\lambda_c$ is the same as the number of $\GP$ singular values associated with the signal. Figure \ref{fig:eig_lambda} shows the distribution of $\hat\lambda_j$ for one of the realizations of the while noise-corrupted signal, in the case of $\bar{n}=4$ and $d=18$. Four pairs of the generalized eigenvalues have a modulus close to 1, while the rest has significantly lower magnitudes. The value of $\lambda_c\approx0.8$ provides a threshold to separate the true and false estimates in this case. Our experience indicates that the optimal value of $\lambda_c\approx0.8$, however it might need to be decreased if the SNR is reduced or the time series length is reduced. We nevertheless propose to keep the value of $\lambda_c$ at the lower end to obtain some false estimates and then discard them based on using the pattern recognitions.

We consider 100 realizations of the time series and estimate frequencies by the ESPRIT and NHSSA. To improve the accuracy of frequency estimation, the size of the auto-covariance matrix for ESPRIT is selected to be $m/3=100$. We consider two realizations of ESPRIT denoted by ESPRIT($\bar{n}$), when $\bar{n}=4$ (an exact number of cosines in the time series) or $\bar{n}=7$ is the assumed number of cosines in the ESPRIT algorithm.

\begin{figure}[ptb]
\centering
\includegraphics[width=.95\columnwidth]{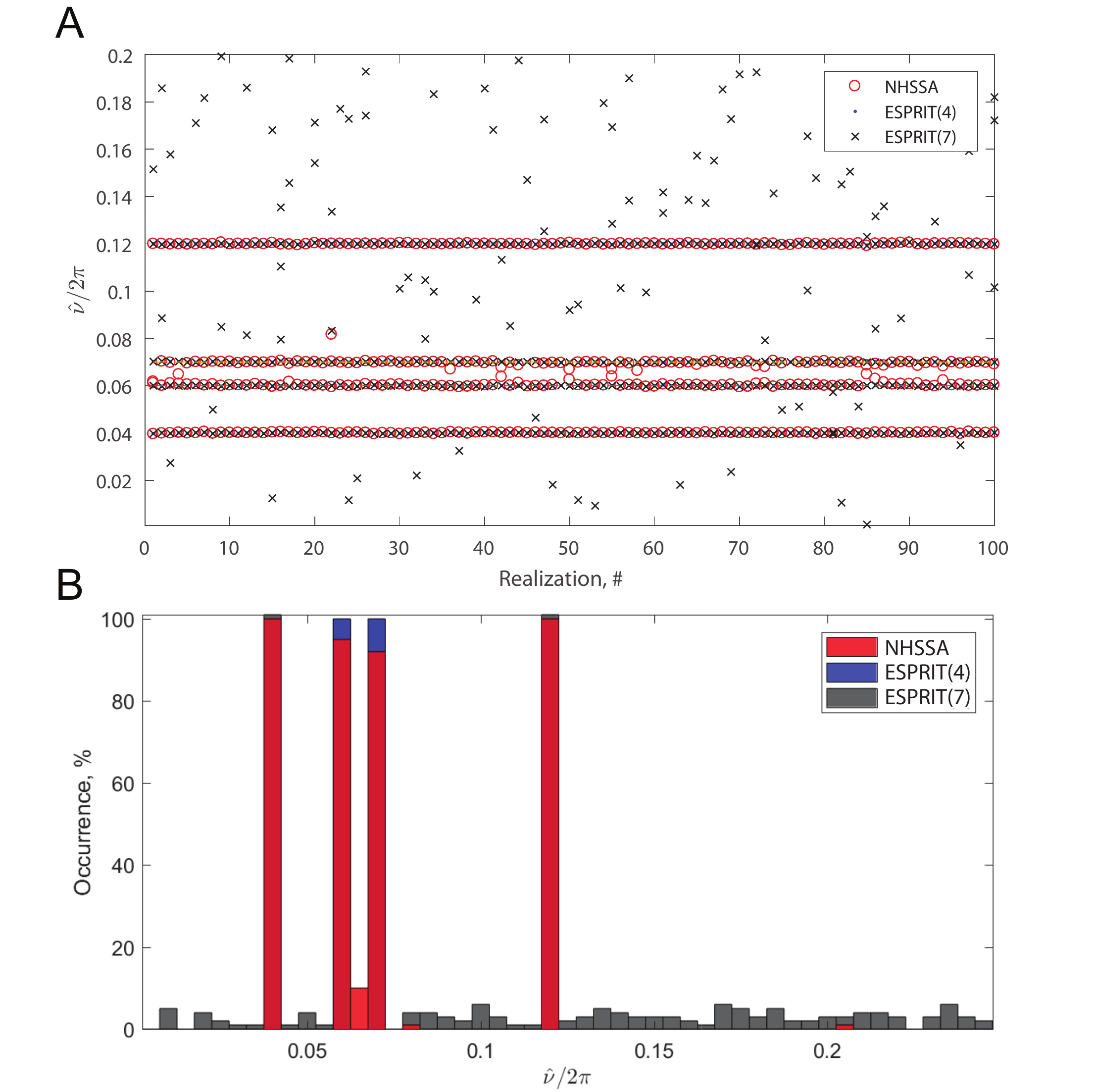}
\caption{(A) Frequencies estimated by NHSSA and ESPRIT for different realizations of the white Gaussian noise. The number of cosines in ESPRIT is assumed to be either 4 or 7. (B) Probability of occurrence for the estimated frequencies.}\label{fig:estimation_wn}
\end{figure}

Figure \ref{fig:estimation_wn}a plots the recovered frequencies for each realization according to each method, i.e. frequencies recovered by NHSSA are plotted by red circles, results of the ESPRIT recovery are shown by dots and crosses. Figure \ref{fig:estimation_wn}b shows the probability of occurrence (a number of times the frequency is recovered within 0.005 intervals, which uniformly span the frequency domain) for the estimated frequencies. Both realizations of ESPRIT almost always recover the frequencies, while NHSSA shows good performance for $\nu/2\pi=0.04$ and $\nu/2\pi=0.12$, while slightly underperforms for $\nu/2\pi=0.06$ and $\nu/2\pi=0.07$. Instead of identifying both of the latter frequencies, NHSSA recovers a single frequency in the middle. We will discuss the separability of frequencies later in this section. Note that all methods recover the frequencies quite well, especially after taking into the account that NHSSA does not have any information about the number of exponentials.

\begin{figure}[pt]
\centering
\includegraphics[width=.95\columnwidth]{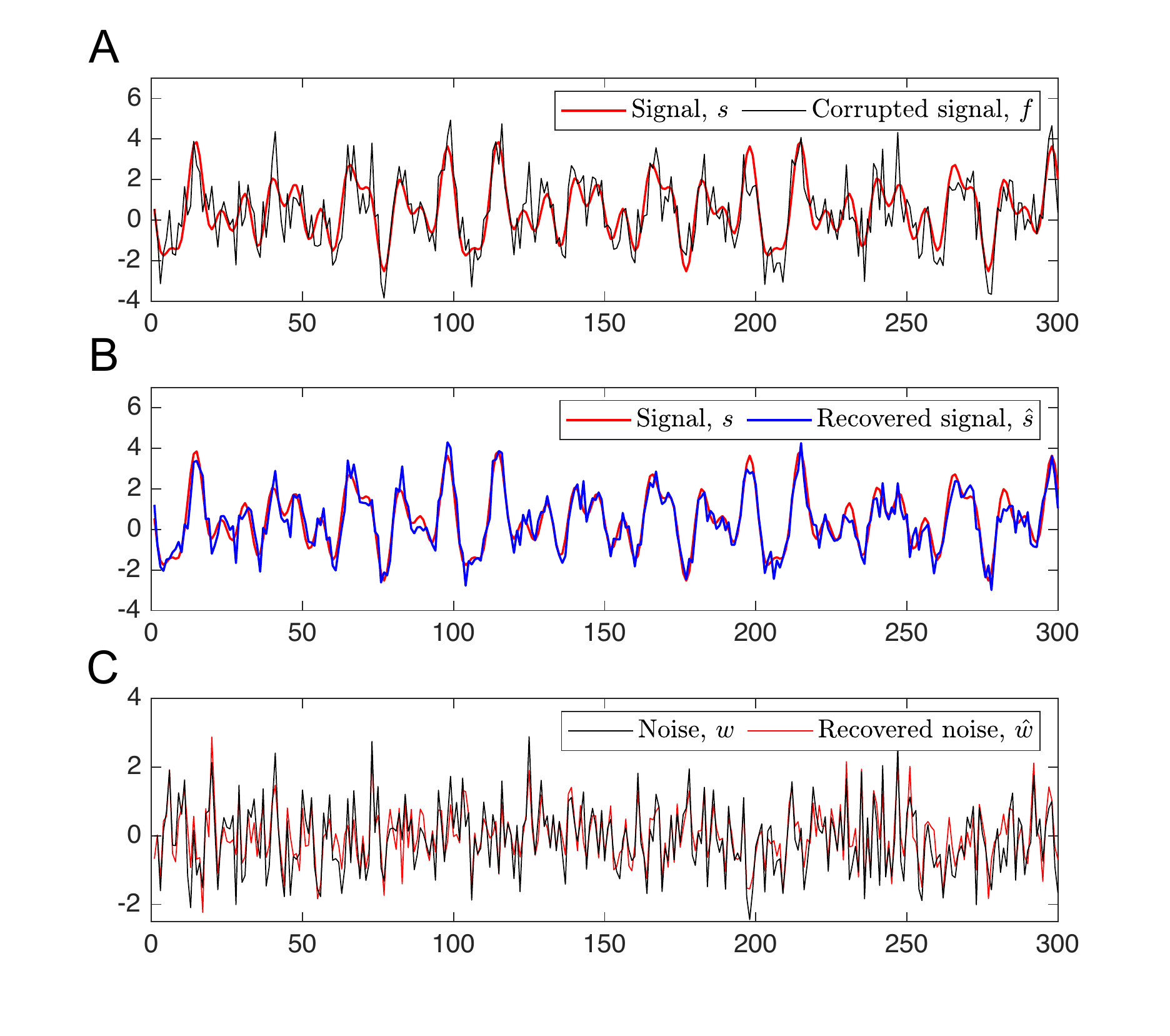}
\caption{(A) Comparison of the original signal $s$ consisting of the four cosines to the corrupted signal $f=s+w$, where $w$ is the white Gaussian noise; (B) Comparison of the original signal $s$ and the recovered ones $\hat{s}$; (C) Comparison of the original noise $w$ and the recovered noise $\hat{w}$.}\label{fig:signal_wn}
\end{figure}

We list the estimated mean and variance for each $\{\hat\nu_k\}_{k=1}^4$ in Table \ref{tab_wn}. The variance of ESPRIT is smaller than that of NHSSA, but the ESPRIT requires a number of cosines as an input variable. Additionally, the ESPRIT with $n=7$ provides additional estimates, shown by black crosses, that are  distributed rather randomly. It is hard to distinguish these false estimates from the true frequencies if only their values are given (as a potential solution one may apply ESPRIT with different sizes of the auto-covariance matrix to investigate whether the estimates are true or false). The NHSSA also has some false estimates, i.e. red circles lying away from the genuine frequencies.

\begin{table}
\caption{Estimation of the frequencies $\{\nu_k\}$ by ESPRIT and NHSSA in the case of time series $s$ corrupted by the white Gaussian or autoregressive noise of the first and second order.}\label{tab_wn} \vskip4mm
\begin{tabular}{|c|c|c|c|c|c|c|} \hline
\multirow{2}{*}{$\nu$}&\multicolumn{3}{|c|}{$\textrm{E}(\hat\nu)$} & \multicolumn{3}{|c|}{$\textrm{Var}(\hat\nu)$}\\\cline{2-7}
&NHSSA&ESPRIT(4) &ESPRIT(7) &NHSSA&ESPRIT(4) &ESPRIT(7) \\\hline
\multicolumn{7}{|l|}{White Gaussian noise, SNR$_{\textrm{dB}}$ = 3.5 dB}\\\hline
0.04&0.040066 &0.040008&0.040018 &4.2092e-08 &2.5401e-08 &2.5111e-08 \\
0.06&0.060517 &0.060011&0.060014 &1.6183e-06 &3.0151e-08 &3.0139e-08 \\
0.07&0.069285 &0.070011&0.070007 &3.1890e-06 &2.1462e-08 &2.2494e-08 \\
0.12&0.119960 &0.119980&0.119970 &4.8482e-08 &2.9076e-08 &3.2675e-08 \\\hline
\multicolumn{7}{|l|}{AR1 noise, SNR$_{\textrm{dB}}$ = 0.9 dB}\\\hline
0.04&0.039971 & 0.039348 & 0.039885 &1.0164e-06& 1.2936e-05 &6.7429e-07 \\
0.06&0.059944 & 0.059894 & 0.059777 &4.8822e-07& 8.3477e-06 &4.3980e-07 \\
0.07&0.069988 & 0.068586 & 0.069860 &3.9226e-07& 8.6346e-06 &2.2843e-07 \\
0.12&0.11999  & 0.116940 & 0.119830 &1.2697e-07& 0.00010443 &8.3213e-08 \\\hline
\multicolumn{7}{|l|}{AR2 noise, SNR$_{\textrm{dB}}$ = 1.5 dB}\\\hline
0.04&0.040022 & 0.040044 & 0.040022 &1.3134e-07& 7.3852e-08 &7.5397e-08 \\
0.06&0.060098 & 0.060107 & 0.060024 &2.8021e-07& 3.1196e-07 &1.0839e-07 \\
0.07&0.069964 & 0.070132 & 0.070059 &3.0026e-07& 1.3255e-07 &1.0902e-07 \\
0.12&0.119770 & 0.120110 & 0.119980 &5.2731e-06& 3.3860e-07 &2.2155e-07 \\\hline
\end{tabular}
\end{table}

Finally, we illustrate separability of the original time series $f=s+w$ into the components, namely: $\hat{s}$ - the recovered signal and $\hat{w}$ - the estimated noise. The recovered signal $\hat{s}$ is obtained by grouping and mapping sequences associated with $\{Z_{k,j}\}_{k=0}^l$, which have a circular phase portrait, to the original space. The noise $\hat{w}$ is estimated by grouping and mapping the rest of $\{Z_{k,j}\}_{k=0}^l$ to the original space as well. Figure \ref{fig:signal_wn}a shows the signal $s$ and the time series $f=s+w$ for one of the realizations of $w$. The recovered signal $\hat{s}$ is compared to the original signal $s$ in Figure \ref{fig:signal_wn}b. Note that the two time series match quite well, but $\hat{s}$ still have some noise. Figure \ref{fig:signal_wn}c shows the comparison of time series for the original $w$ and estimated $\hat{w}$ noise, which match quite good as well. We emphasize that the decomposition of the time series $f=\hat{s}+\hat{w}+O(\epsilon)$ is obtained straight from grouping appropriate $\{Z_{k,j}\}_{k=0}^l$ and mapping them to the original space. The proposed method thus allows us to de-noise the original time series. Despite some inspiring empirical observations, further investigations are required to derive sharp estimates for $s-\hat{s}$ and $w-\hat{w}$.

\subsection{Signal corrupted by the autoregressive noise}
To further test the proposed method, we consider a signal $s$ consisting of the same four cosines, but now it is corrupted by the noise $w$, which is generated according to an autoregressive (AR) process of either the first or second order:
\[
w(k)=0.7w(k-1)+\xi(k),
\]
or
\[
w(k)=0.7w(k-1)-0.4w(k-2)+\xi(k).
\]
Here, $\xi(k)$ is the white Gaussian noise such that $\textrm{E}(\xi)=0$ and $\textrm{Var}(\xi)=1$; the SNR$_{\textrm{dB}}$ for the former and latter models is 0.6 and 1.5 dB, respectively.

\begin{figure}[H]
\centering
\includegraphics[width=.95\columnwidth]{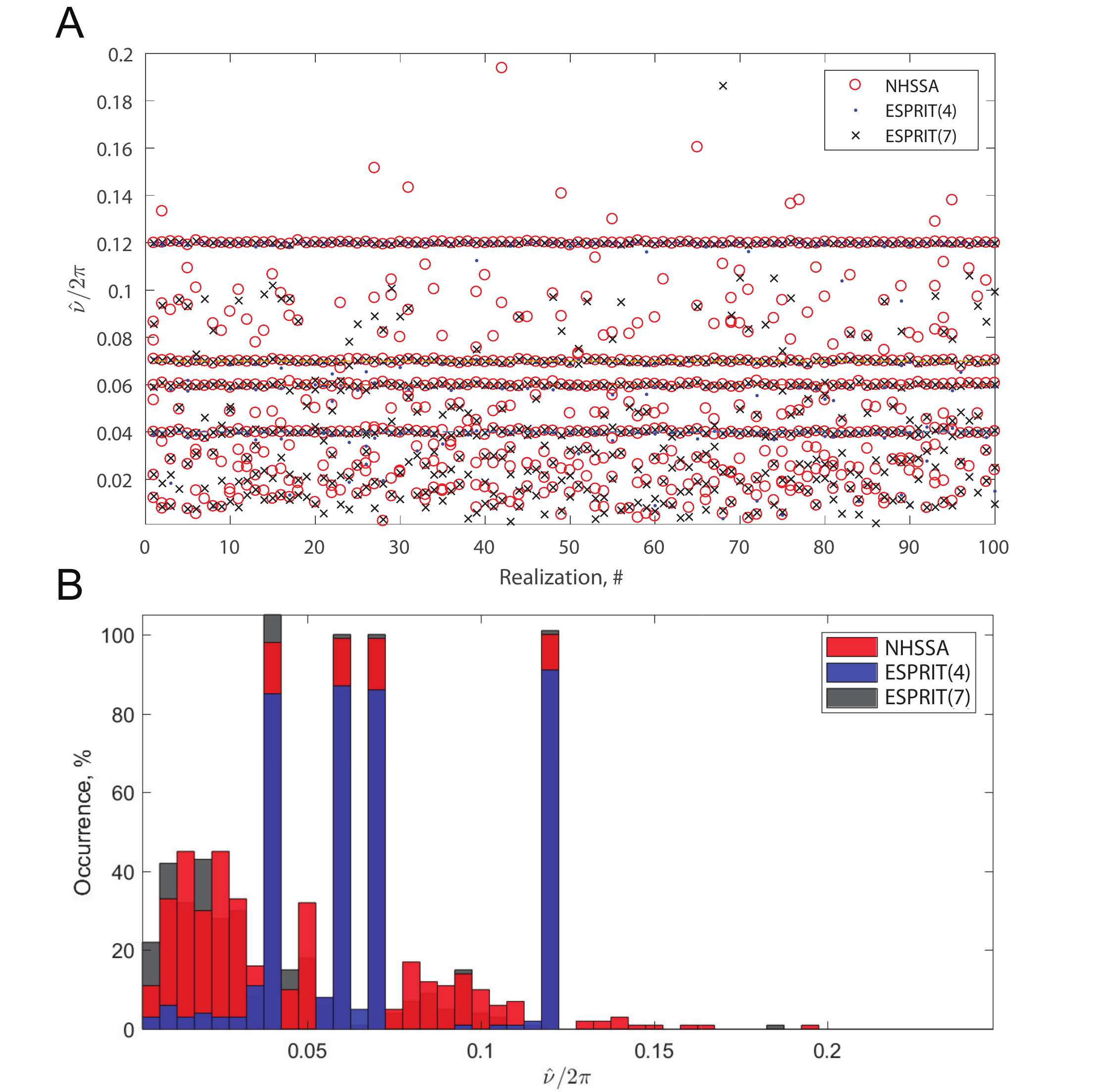}
\caption{(A) Frequencies estimated by NHSSA and ESPRIT for different realizations of the autoregressive noise of the first order. The number of cosines in ESPRIT is assumed either to be 4 or 7; (B) Probability of occurrence for the estimated frequencies.}\label{fig:estimation_ar1}
\end{figure}

\begin{figure}[H]
\centering
\includegraphics[width=.95\columnwidth]{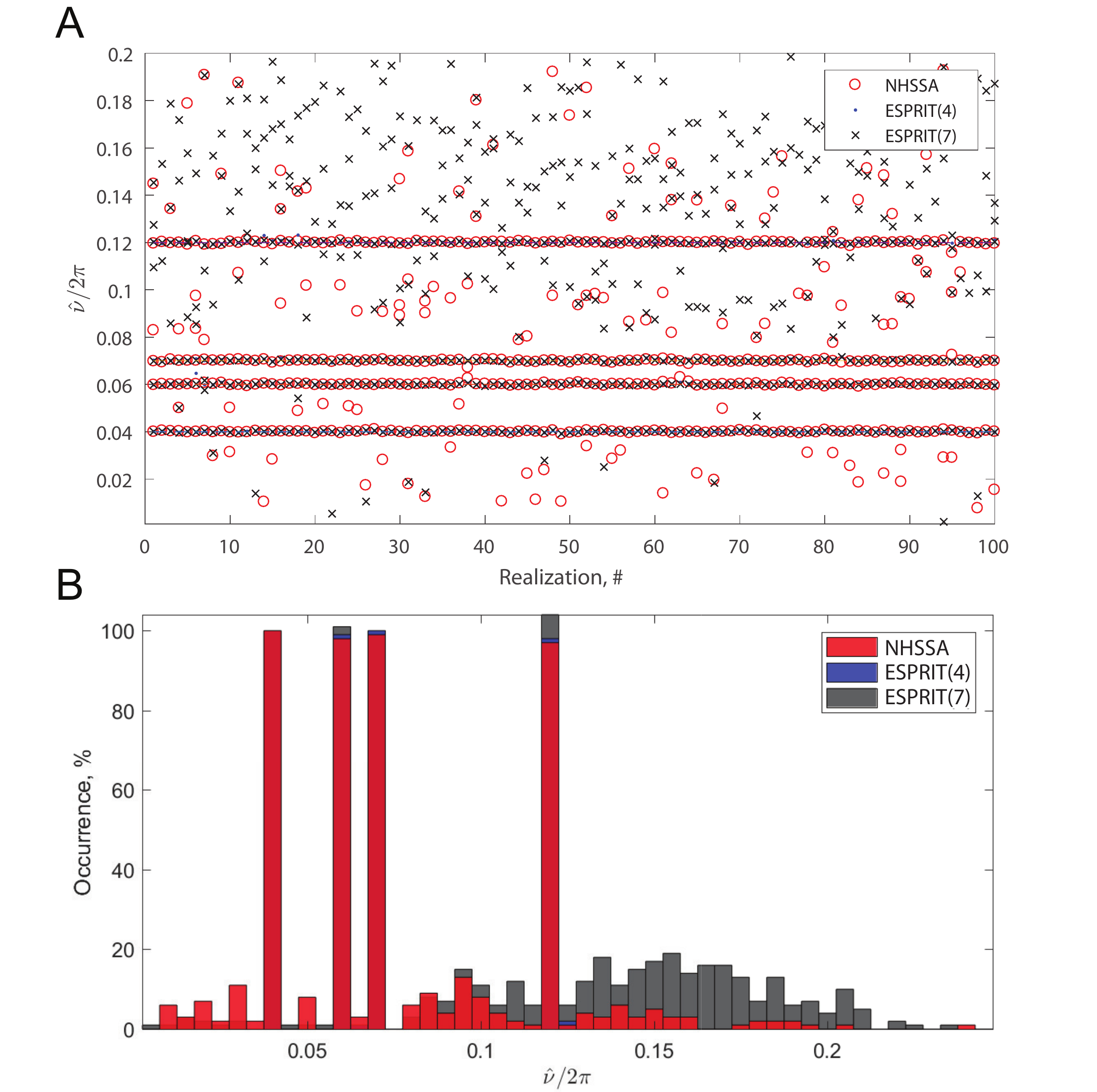}
\caption{(A) Frequencies estimated by NHSSA and ESPRIT for different realizations of the autoregressive noise of the second order. The number of cosines in ESPRIT is assumed either to be 4 or 7; (B) Probability of occurrence for the estimated frequencies.}\label{fig:estimation_ar2}
\end{figure}

Note that for the AR noise, the eigenvalues of $\Gamma_0$ do not have a clear jump as shown in Figure \ref{fig:eig} and the number of exponentials for ESPRIT could be estimated between $j=9$ and $j=15$ or higher (since each cosine is associated with two exponentials, and thus the number of cosines could be between 4 and 7). After computing the condition number $\textrm{cond}(\XS)$ on the grid of $(d,\bar{m})$, we choose $\bar{m}=3$ and $d=18$. As before, we consider 100 different realizations of the noise and estimate frequencies by the same three methods. Results of the frequency recovery for the AR1 and AR2 noise models are shown in Figures \ref{fig:estimation_ar1} and \ref{fig:estimation_ar2}, respectively. Note that AR1 noise model generates a significant number of false estimates for ESPRIT, there is also a comparable number of false estimates for NHSSA, but the latter could be discarded using the pattern recognitions. For the AR2 noise model, NHSSA performs significantly better than ESPRIT(7), but less effective than ESPRIT(4). However, ESPRIT(4) has an advantage by exploiting information regarding the correct number of exponentials in the signal. If the number of exponentials is increased, e.g. as in ESPRIT(7), false estimates occur. For the sake of completion we list the mean and variance for each recovered frequency in Table \ref{tab_wn}.

\subsection{Separability}
\begin{figure}[H]
\centering
\includegraphics[width=.95\columnwidth]{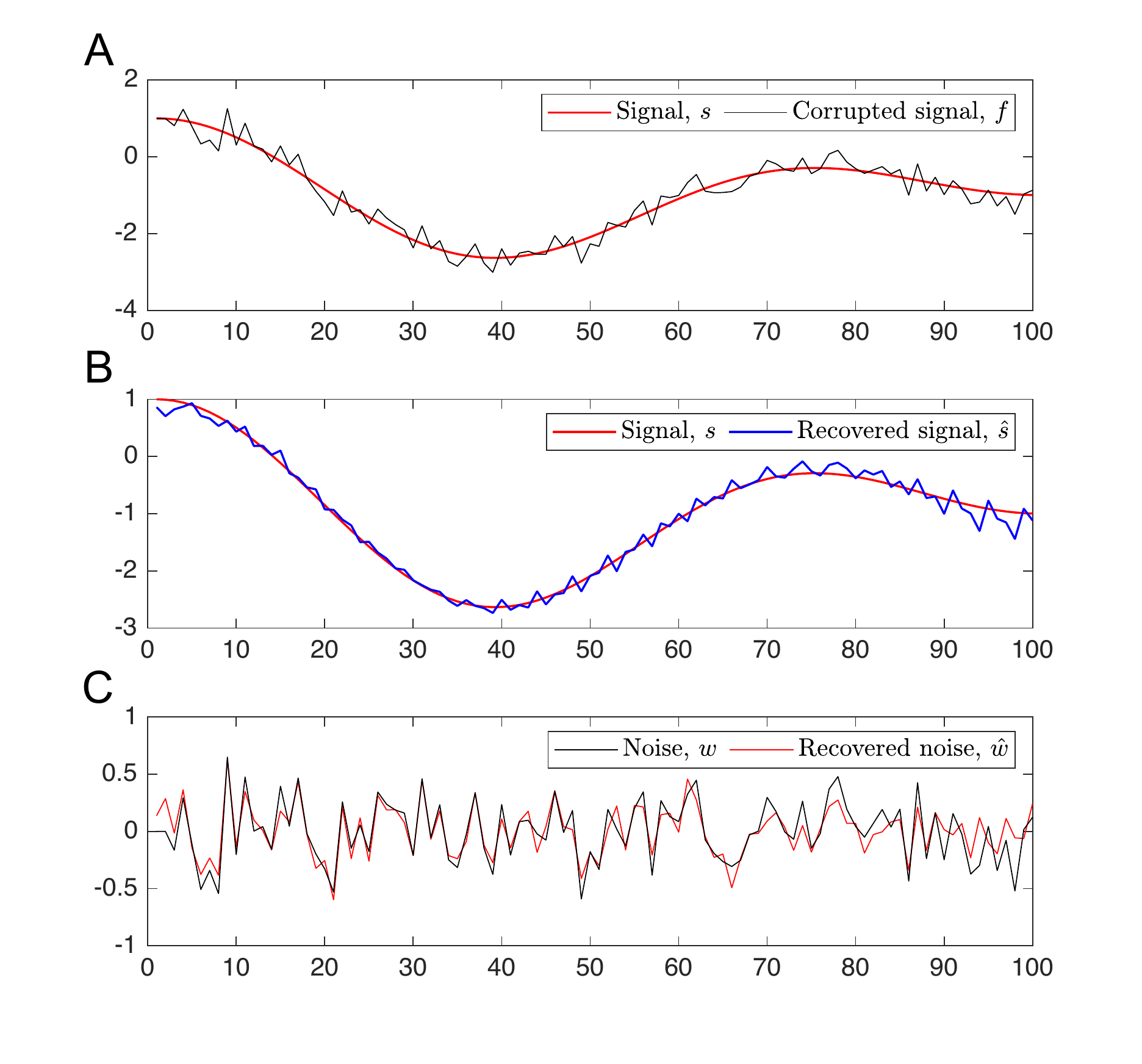}
\caption{(A) Comparison of the original signal $s$ consisting of the two cosines with closely lying frequencies to the corrupted signal $f=s+w$, where $w$ is the white Gaussian noise; (B) Comparison of the original signal $s$ and the recovered ones $\hat{s}$; (C) Comparison of the original noise $w$ and the recovered noise $\hat{w}$.}\label{fig:signal_together}
\end{figure}

In this section, we consider a signal composed of two cosines with close frequencies that are hard to detect simultaneously due to the shortness of the time series $f$, i.e. the sum of cosines is sampled at $m=100$ points. Namely, we consider $\nu_1/2\pi=0.01$ and $\nu_2/2\pi=0.015$, $c_k=c_{-k}=0.5$ for $k=1,2$, and $c_0=-1$; the noise $w$ is assumed white Gaussian with the zero mean $\textrm{E}(w)=0$ and $\textrm{Var}(w)=1/16$. The signal $s$ and time series $f$ are shown in Figure \ref{fig:signal_together}a; the SNR is 15 dB.

\begin{figure}[H]
\centering
\includegraphics[width=.95\columnwidth]{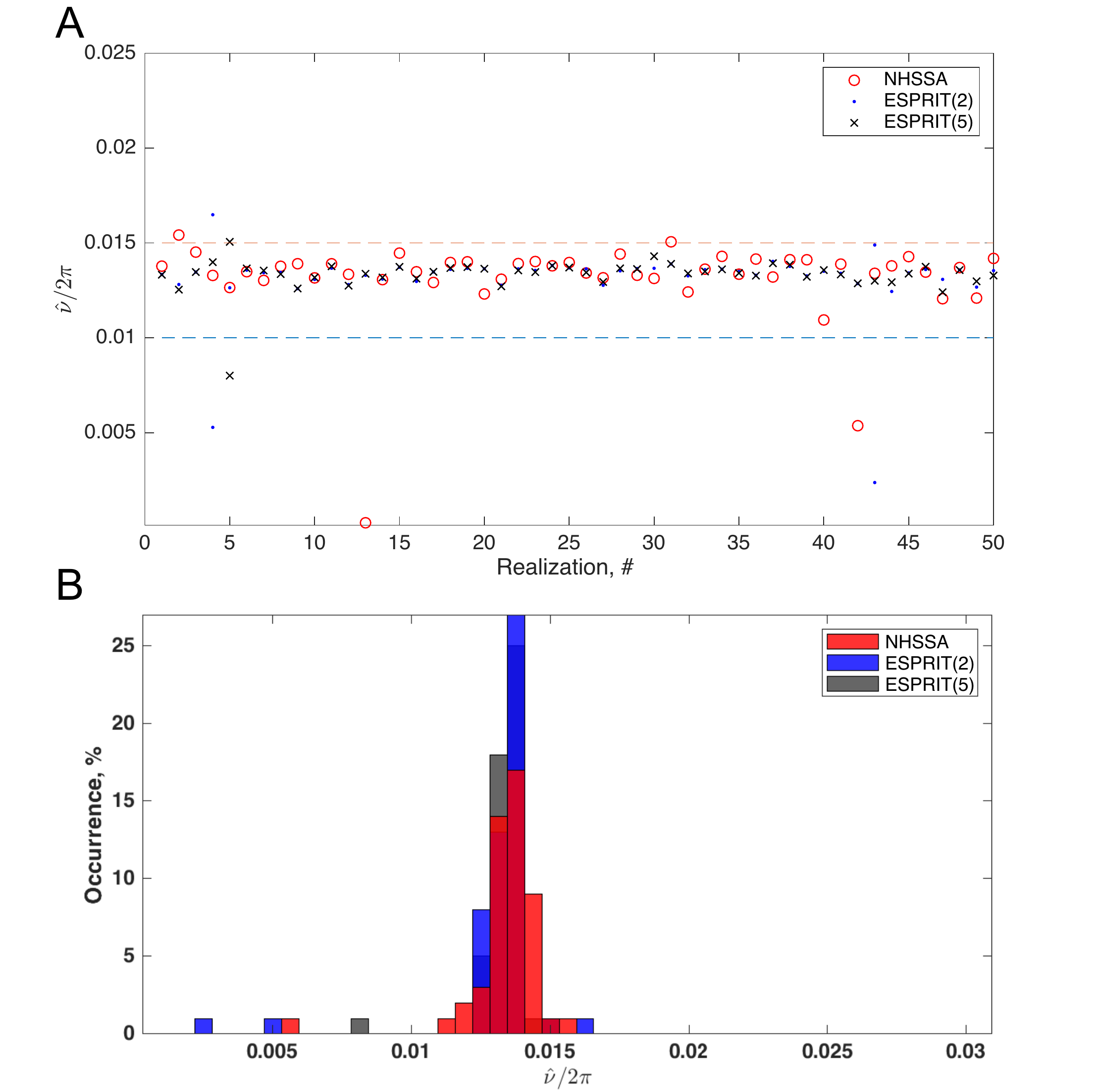}
\caption{(A) Frequencies estimated by NHSSA and ESPRIT for different realizations of the white Gaussian noise. The number of cosines in ESPRIT is assumed to be either 2 or 5; (B) Probability of occurrence for the estimated frequencies.}\label{fig:estimation_together}
\end{figure}

\begin{figure}[H]
\centering
\includegraphics[width=.95\columnwidth]{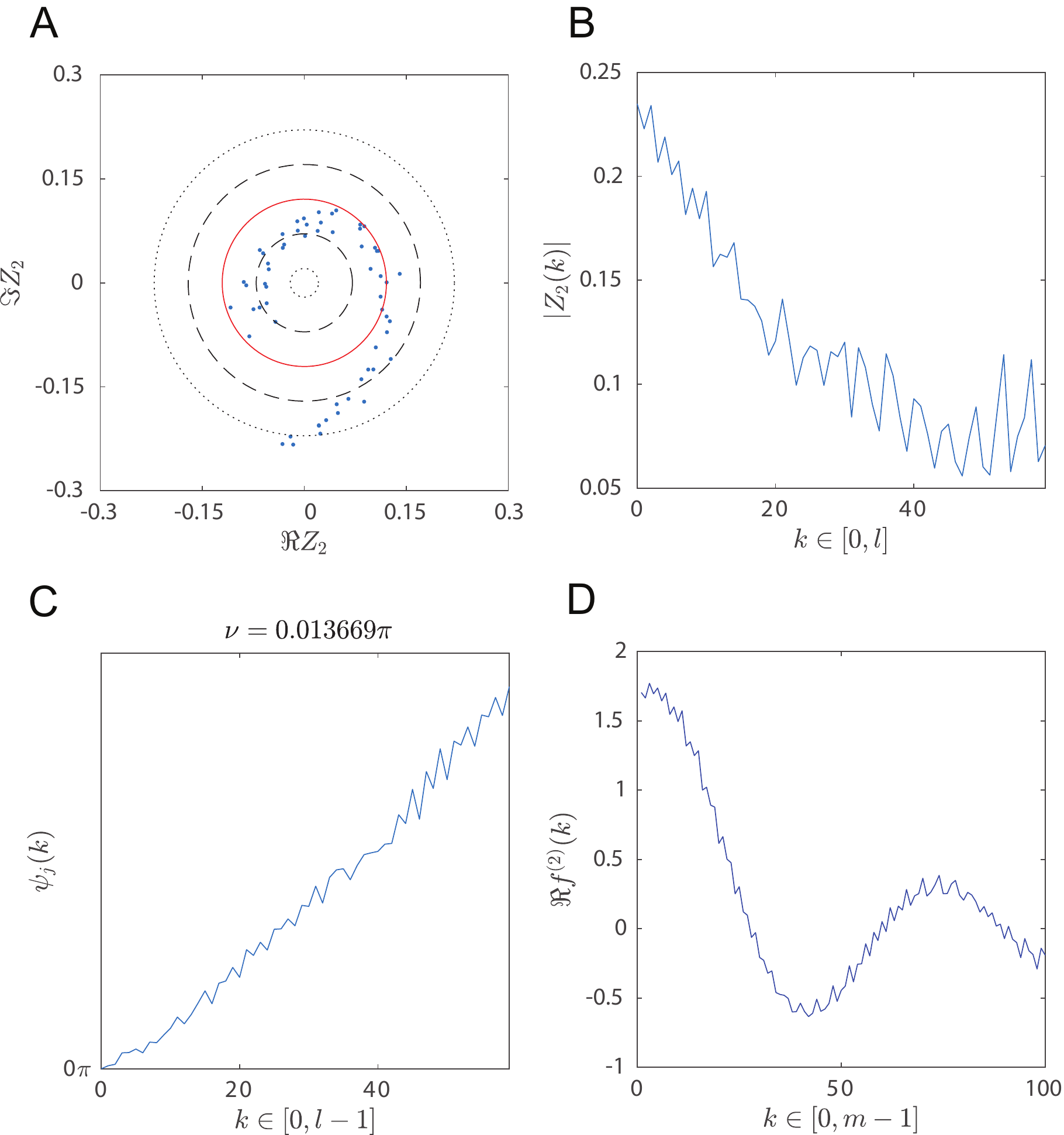}
\caption{Typical results for the sequence $\{Z_{k,j}\}_{k=0}^l$ associated with an average of two exponentials with closely lying frequencies. (A) The phase portrait $\{(\Re(Z_{k,j}),\Im(Z_{k,j}))\}_{k=0}^l$ is a spiral; dashed and dotted lines represent one and two standard deviations from the mean modulus; (B) The modulus of $\{Z_{k,j}\}_{k=0}^l$ is generally decreasing; (C) The phase $\psi_j(k)$ is a linearly increasing function; (D) Mapping of $\{Z_{k,j}\}_{k=0}^l$ back to the original space $f^{(j)}(k)$ is a damping cosine.}\label{fig:z2_together}
\end{figure}

Estimation of the frequencies by both ESPRIT and NHSSA is shown in Figure \ref{fig:estimation_together}. Note that ESPRIT(2) with the correct number of cosines fails to estimate both frequencies simultaneously, and rather estimates a single frequency somewhere in the middle between $\nu_1/2\pi$ and $\nu_2/2\pi$. The NHSSA also fails to distinguish both cosines and recovers a single frequency - some sort of averaging of $\nu_1/2\pi$ and $\nu_2/2\pi$ as well. This frequency is related to an eigenvalue pair (the cosine consists of two exponentials) lying close to the unit circle, while other eigenvalues have significantly lower moduli and are related to noise. The phase portrait $\{(\Re(Z_{k,j}),\Im(Z_{k,j}))\}_{k=0}^l$ of the sequence $\{Z_{k,j}\}_{k=0}^l$ associated with the pair lying close to the unit circle is a spiral, see Figure \ref{fig:z2_together}a.

We stress that the original problem of exponential recovery was concerned with finding frequencies $\{\nu_k: {\nu_k}\in\mathcal{R}\}$. However, both ESPRIT \cite{Kundu95} and NHSSA could be applied to recover damping exponentials with frequencies $\{\nu_k:{\nu_k}\in\mathcal{C}_+\}$, where $\mathcal{C}_+$ is the upper complex plane, which discussion is beyond the scope of this manuscript. Therefore, instead of recovering both exponentials with closely lying frequencies, the NHSSA approximated them by a single damping cosine, as could be observed in the mapping of $\{Z_{k,j}\}_{k=0}^l$ back to the original space, see Figure \ref{fig:z2_together}d. This information is not revealed by standard ESPRIT, since it only relies on the eigenvalues and discards information carried by the eigenvectors. The noise $\hat{w}$ and $\hat{s}$ are both recovered in NHSSA by grouping appropriate eigenvalues and mapping back to the original space, see Figures \ref{fig:signal_together}b and \ref{fig:signal_together}c. Nevertheless, further investigations are necessary to understand when the recovery of two exponentials sampled on a short interval is feasible.


\section{Conclusions}
We present a new method of estimating exponentials and their frequencies in the time series. The proposed method decomposes the time series consisting of several exponentials into components by casting the information vectors into a new basis. Each component corresponds either to only one of the exponentials or to noise. For the information-carrying component one of many frequency estimation techniques (e.g. ESPRIT, MUSIC, ML, etc.) could be applied to recover a single exponential. The overall accuracy of the proposed method is comparable with that of the widely used ESPRIT method, if the latter is provided with the number of exponentials in the signal. Furthermore, when the model order (number of exponentials) in ESPRIT is overestimated, the proposed method can reduce the number of false frequency estimates, as shown by numerical examples.

One of the significant benefits of the proposed approach is a way to distinguish false and true frequency estimates using the pattern recognition. The primary automatization of the pattern recognition is completed by discarding noise-related components, associated with the eigenvectors that have a modulus less than a certain threshold $\lambda_c$. At the second stage, the phase portrait and phase dynamics for the remaining components could be visually analyzed. Images associated with the true frequencies have phase portraits resembling unit circles and phase dynamics with zero or a minimal number of phase wrapping events. Closeness of the phase portrait to the unit circle and mapping of the component back to the original space could provide certain levels of confidence that the component is associated with the exponential or noise. False frequencies associated with the phase portraits that have a random structure. At the same time further research is needed to produce a fully automatic pattern recognition algorithm.

Finally, we note that under certain conditions the proposed method could be generalized to estimate exponentials in the time series where some measurements are missing \cite{Avdonin09}.

\section{Acknowledgements}
We would like to thank A. Rybkin and V. Romanovsky for all their advice, patience, critique and reassurances along the way. We also appreciate comments and suggestions by the anonymous reviewer, who greatly helped to improve the quality of this manuscript. This research was funded by ARCSS Program and by the Polar Earth Science Program, Office of Polar Programs, National Science Foundation and by the State of Alaska.

\appendix
\section{Derivations of main mathematical results}\label{ap:NoiseVariance}

\textbf{Proof of Formula \ref{lem12}:}
Note that the cost function $J$ defined by (\ref{eq:11}) has the
following representation
\[
J(\mathbf{A})=\mathbf{tr}({\summa}E_kE_k^*)=\mathbf{tr}\big({\summa}(Y_{k+1}{-}\mathbf{A}Y_k)(Y_{k+1}{-}\mathbf{A}Y_k)^*\big).
\]
Considering the change of variables
$\mathbf{A}{=}\mathbf{\Gamma}_1\mathbf{\Gamma}_0^{-1}{+}\mathbf{U}$,
it is easy to see that the term corresponded to the first power of
$\bigU$ vanishes by the definition of $\mathbf{\Gamma}_1$ and
$\mathbf{\Gamma}_0$; and the cost function is
\begin{equation}
J(\mathbf{A}){=}\frac1l\mathbf{tr}\big\{(\XS{-}\mathbf{\Gamma}_1\mathbf{\Gamma}_0^{-1}\XP)(\XS{-}\mathbf{\Gamma}_1\mathbf{\Gamma}_0^{-1}\XP)^{*}\big\}{+}\frac1l\mathbf{tr}\big\{\mathbf{U}\XP\XP^{*}\mathbf{U}^{*}\big\}.
\end{equation}
Since the matrix
$\mathbf{\Gamma}_0=\mathbf{W}\mathbf{\Lambda}\mathbf{W}^*$ is
non-negative definite, the matrix of its eigenvalues
$\mathbf{\Lambda}=diag(\lambda_1,\dots,\lambda_d)$ has only
non-negative entries. Therefore, introducing
$\mathbf{\tilde{U}}=\mathbf{U}\mathbf{W}$, we have
\[
\mathbf{tr}\big(\mathbf{U}\mathbf{\Gamma}_0\mathbf{U}^{*}\big){=}\mathbf{tr}\big(\tilde{\bigU}\mathbf{\Lambda}\tilde{\bigU}^{*}\big){=}
\sum_{i=1}^d\lambda_i(\sum_{j=1}^d{|\tilde{U}_{ij}|^2}){\ge}0,
\]
and hence the unique extremum exists at $\mathbf{U}=0$, or
$\mathbf{A}=\mathbf{\Gamma}_1\mathbf{\Gamma}_0^{-1}$.

\textbf{Proof of Formula \ref{lem21}:}
Let the Jordan decomposition of $\mathbf{\Omega}$ is
\begin{equation}\label{eq:jor}
\mathbf{\hat{V}^{-1}}\mathbf{\hat\Omega}=\mathbf{\hat\Lambda}\mathbf{\hat{V}^{-1}},
\end{equation}
where $\mathbf{V}$ and $\mathbf{\Lambda}$ are matrices consisting
the eigenvectors and eigenvalues of $\mathbf{\hat\Omega}$,
respectively. Then after multiplying (\ref{eq:jor}) by the matrix
$\mathbf{V}$ from the left and exploiting the definition
(\ref{lem21}) of $\mathbf{\hat{V}}$, we have
\begin{equation}\label{eq:repR}
\mathbf{V}^{-1}\mathbf{\hat\Omega}\mathbf{V}+\epsilon\mathbf{R}\mathbf{V}^{-1}\mathbf{\hat\Omega}\mathbf{V}=\mathbf{\hat\Lambda}+\epsilon\mathbf{\hat\Lambda}\mathbf{R}.
\end{equation}
Note that since matrices $\mathbf{\Gamma}_0$ and $\mathbf{\Gamma}_1$ are second degree polynomials with respect to $\epsilon$, the Taylor series expansion of
$\mathbf{\Omega}=\mathbf{\Gamma}_1\mathbf{\Gamma}_0^{-1}$ has all powers of $\epsilon$:
\[
\mathbf{\hat\Omega}=\mathbf{\Omega}+\sum_{m=1}^\infty\epsilon^m\mathbf{\Omega}^{(m)},
\]
and $\mathbf{\hat\Omega}$ is an analytic function of $\epsilon$.
From the theory of Linear Algebra it is known that the eigenvalues
$\mathbf{\hat\Lambda}$ of $\mathbf{\hat\Omega}$ are analytic
functions of $\epsilon$ and are given by
\[
\mathbf{\hat\Lambda}=\mathbf{\Lambda}+\sum_{m=1}^\infty\epsilon^m\mathbf{\Lambda}^{(m)},
\]
where $\mathbf{\hat\Lambda}$ and $\mathbf{\Lambda}$ are the diagonal matrices containing eigenvalues of $\mathbf{\Omega}$ and $\mathbf{\hat\Omega}$, respectively. By the same argument, it is possible to prove that $\mathbf{\hat{V}}$ and hence $\mathbf{\hat{R}}$ are analytic matrices too. Therefore, the representation:
\[
\mathbf{R}(\epsilon)=\sum_{n=0}^\infty\epsilon^n\mathbf{R}^{(n+1)}
\]
holds. Substituting this formula into (\ref{eq:repR}) and equating the terms in front of the same powers of $\epsilon$, we derive
\[
\mathbf{R}^{(n)}\mathbf{\Lambda}-\mathbf{\Lambda}\mathbf{R}^{(n)}=\mathbf{\Lambda}^{(n)}+\mathbf{G}^{(n)},
\]
where
$\mathbf{G}^{(n)}=-\mathbf{V}^{-1}\mathbf{\Omega}^{(n)}\mathbf{V}+\sum_{k=1}^{n-1}\left(\mathbf{\Lambda}^{(k)}\mathbf{R}^{(n-k)}-\mathbf{R}^{(k)}\mathbf{V}^{-1}\mathbf{\Omega}^{(n-k)}\mathbf{V}\right)$.
Therefore
\[
\mathbf{\Lambda}^{(n)}=-\textrm{diag}\left(\mathbf{G}^{(n)}\right), \ \ \ \
\mathbf{R}^{(n)}_{ij}=\left\{\begin{array}{cc}
-\frac{G^{(n)}_{ij}}{\lambda_i-\lambda_j} & i{\ne}j \\
0 & i{=}j
\end{array}\right..
\]


\end{document}